\documentclass[10pt,journal,letterpaper,compsoc]{IEEEtran}

\usepackage{url}
\usepackage{amsmath}
\usepackage{amssymb}

\usepackage{graphicx}
\usepackage{placeins}
\usepackage[draft,footnote,nomargin]{fixme}
\usepackage[nolist]{acronym}
\begin{acronym}[]
\acro{ACK}{Acknowledgment}
\acro{AMR}{Adaptive Multi-Rate}
\acro{AP}{Access Point}
\acro{ARQ}{Automatic Repeat-reQuest}
\acro{ARF}{Auto Rate Fallback}
\acro{API}{Application Programming Interface}
\acro{BER}{Bit Error Rate}
\acro{BS}{Base Station}
\acro{BSS}{Basic Service Set}
\acro{CFP}{Contention Freeze Periode}
\acro{CPU}{Central Processing Unit}
\acro{CSMA/CA}{Carrier Sense Multiple Access with Collision Avoidance}
\acro{CW}{Contention Window}
\acro{CWN}{Cooperative Wireless Network}
\acro{CTS}{Clear to Send}
\acro{DCF}{Distributed Coordination Function}
\acro{DIFS}{Distributed Interframe Space}
\acro{ERP}{Extended Rate PHY}
\acro{FEC}{Forward Error Correction}
\acro{FCS}{Frame Check Sequence}
\acro{FF}{Finite Field}
\acro{GF}{Galois Field}
\acro{GPS}{Global Positioning System}
\acro{HTTP}{Hyper Text Transfer Protocol}
\acro{IBSS}{Independent BSS}
\acro{IC}{Interference Cancellation}
\acro{ICST}{Institute for Computer Sciences, Social-Informatics and Telecommunications Engineering}
\acro{IFS}{Interframe Space}
\acro{IP}{Internet Protocol}
\acro{IPTV}{Internet Protocol TeleVision}
\acro{JCN}{Journal of Communications and Networks}
\acro{KL}{Kullback-Leibler}
\acro{LAN}{Local Area Network}
\acro{LNC}{Linear Network Coding}
\acro{LOS}{Line Of Sight}
\acro{MAC}{Medium Access Control}
\acro{MANET}{Mobile Ad hoc NETwork}
\acro{MIMO}{Multiple Input Multiple Output}
\acro{MIT}{Massachusetts Institute of Technology}
\acro{MTBF}{Mean Time Between Failure}
\acro{NAV}{Network Allocation Vector}
\acro{NEP}{Nokia Energy Profiler}
\acro{NC}{Network Coding}
\acro{NLOS}{Non Line Of Sight}
\acro{NIC}{Network Interface Card}
\acro{OSI}{Open Systems Interconnect}
\acro{PC}{Personal Computer}
\acro{PDA}{Personal Digital Assistant}
\acro{PEP}{Packet Error Probability}
\acro{PHY}{Physical Layer}
\acro{PLCP}{Physical Layer Convergence Procedure}
\acro{PNC}{Physical layer Network Coding}
\acro{PPDU}{PLCP Protocol Data Unit}
\acro{QoS}{Quality of Service}
\acro{RLNC}{Random Linear Network Coding}
\acro{RTS}{Request to Send}
\acro{SAIC}{Single Antenna Interference Cancellation}
\acro{SIMD}{Single Instruction Multiple Data}
\acro{SNMP}{Simple Network Managment Protocol}
\acro{SNR}{Signal-to-Noise Ratio}
\acro{SIFS}{Short Interframe Space}
\acro{SRLNC}{Sparse Random Linear Network Coding}
\acro{SSE}{Streaming SIMD Extensions}
\acro{SSID}{Service Set Identifier}
\acro{UDP}{User Datagram Protocol}
\acro{UI}{User Interface}
\acro{VANET}{Vehicular Ad-hoc Network}
\acro{VoIP}{Voice over Internet Protocol}
\acro{WBN}{Wireless Broadcast Network}
\acro{WLAN}{Wireless Local Area Network}
\acro{WMN}{Wireless Mesh Network}
\acro{pmf}{Probability Mass Function}
\acro{TBTT}{Target Beacon Transmision Times}
\acro{TCP}{Transmission Control Protocol}
\end{acronym}

\usepackage{subfig}
\usepackage{multirow}
\usepackage{booktabs}

\usepackage{color,colortbl}
\definecolor{myorange}{rgb}{.8886, 0.5782, 0.2645}
\definecolor{mygreen}{rgb}{0.7098, 0.7961, 0.5916}

\usepackage{tikz}
\usetikzlibrary{matrix,arrows,decorations.pathreplacing,decorations.markings}

\newcommand{\scaletikz}[0]{1.4}

\usepackage[vlined,ruled,linesnumbered]{algorithm2e}
\usepackage{algpseudocode} 
\SetAlgoSkip{}
\renewcommand{\algref}[1]{Algorithm~\ref{#1}}

\usepackage{cite}

\newcommand{\vect}[1]{\boldsymbol{\lowercase{#1}}}

\newcommand{\vecttilde}[1]{\vect{\tilde{#1}}}

\newcommand{\mat}[1]{\boldsymbol{\uppercase{#1}}}
\newcommand{\mathat}[1]{\hat{\mat{#1}}}

\newcommand{\field}[1]{\boldsymbol{ \mathbb{#1}} }

\newcommand{\secref}[1]{Section~\ref{sec:#1}}

\newcommand{\figref}[1]{\figurename~\ref{fig:#1}}

\newcommand{\tabref}[1]{Table~\ref{tab:#1}}

\renewcommand{\eqref}[1]{Equation~(\ref{eq:#1})}

\title{Perpetual Codes for Network Coding}

\author{
\IEEEauthorblockN{Janus Heide\IEEEauthorrefmark{1}, Morten V. Pedersen\IEEEauthorrefmark{1}, Frank H.P. Fitzek\IEEEauthorrefmark{1} and Muriel M\'{e}dard}\IEEEauthorrefmark{2}\\
\IEEEauthorblockA{
\IEEEauthorrefmark{1}
Aalborg University, Aalborg, Denmark,
Email: [jah$|$mvp$|$ff]@es.aau.dk\\
\IEEEauthorrefmark{2}
Massachusetts Institute of Technology,
Cambridge, Massachusetts,
Email: medard@mit.edu\\
}}

\begin{document}

\IEEEcompsoctitleabstractindextext{%
\begin{abstract}
\ac{RLNC} provides a theoretically efficient method for coding. Some of its practical drawbacks are the complexity of decoding and the overhead due to the coding vectors.
For computationally weak and battery-driven platforms, these challenges are particular important. 
In this work, we consider the coding variant \textit{Perpetual codes} which are sparse, non-uniform and the coding vectors have a compact representation. The sparsity allows for fast encoding and decoding, and the non-uniform protection of symbols enables recoding where the produced symbols are indistinguishable from those encoded at the source.
 The presented results show that the approach can provide a coding overhead arbitrarily close to that of \ac{RLNC}, but at reduced computational load. The achieved gain over \ac{RLNC} grows with the generation size, and both encoding and decoding throughput is approximately one order of magnitude higher compared to \ac{RLNC} at a generation size of 2048.
Additionally, the approach allows for easy adjustment between coding throughput and code overhead, which makes it suitable for a broad range of platforms and applications.
%

%
%

\end{abstract}


}

\maketitle
\IEEEdisplaynotcompsoctitleabstractindextext
\IEEEpeerreviewmaketitle


\section{Introduction}



\ac{NC} is a promising paradigm~\cite{ahlswede} that has been shown to provide benefits in many different networks and applications.
\ac{NC} enables coding at individual nodes in a communication network, and thus is fundamentally different from the end-to-end approach of channel and source coding. With \ac{NC}, packets are no longer treated as atomic entities since they can be combined and re-combined at any node in the network.
This allows for a less restricted view on the flow of information in networks, which can be particularly helpful when building distribution systems for less structured networks such as meshed, peer-to-peer or highly mobile networks.

In this work, we focus on random \ac{NC} approaches, i.e. \ac{RLNC}~\cite{rlnc}, and disregard deterministic coding. The reason is that our primary interest is cooperative and highly mobile wireless networks, which fit perfectly with the highly decentralized nature of \ac{RLNC}. In particular, \ac{RLNC} reduces the signaling overhead and increases robustness towards changing channel conditions in the network. At the same time, it allows for the construction of much simpler distribution systems, which is desirable from an engineering point of view.

Unfortunately, \ac{RLNC} is inherently computationally demanding that has spawned several efforts to produce optimized implementations and modify the underlying code~\cite{Yang:2011:LFT:2093698.2093815,NCRealWorld}. Even though several solutions and implementations have been declared to provide \textit{sufficient coding throughput} continued efforts are valid as they can ensure higher coding throughput.
Computational resources can be conserved tasks, such as video decoding, and the energy consumption introduced by coding can be reduced further. This is of particular importance when \ac{NC} is deployed on battery-driven devices with modest computational capabilities.

This paper presents our work on applying Perpetual Codes, which was suggested and named in the unpublished draft~\cite{perpetual}, for \ac{NC} systems.
The encoding is sparse and non-uniform which allows for fast decoding as \textit{fill-in}~\cite{Ingram_minimumdegree} is avoided while recoding is still possible.
The approach presented here is similar to what is called a \textit{smooth perpetual code} in~\cite{perpetual}, but with two significant differences, neither zero padding nor a pre-code is used. This simplifies the analysis, but complicates the final decoding step. 
We describe how encoding and decoding can be performed and analyze the overhead and the complexity. We verify our results with our own C++ implementation, which also provides practical throughput results. Furthermore, we describe, implement, and evaluate recoding which was not considered in~\cite{perpetual}.
The main insight from our results is that \ac{RLNC} is a better choice at low to medium generation sizes, but perpetual codes are more suitable at medium to high sized generations. Hence, perpetual codes are not a substitution, but a supplement to \ac{RLNC}.



%
%





This paper is primarily intended for researchers and developers who work with reliable data distribution on wireless and mobile platforms. Therefore we provide a short overview of \ac{RLNC} and related work in \secref{related}. The approach to encoding, decoding and recoding is presented in \secref{code} together with algorithms aimed at implementations in C or C++. \secref{analysis} provides analysis of the performance of the code in terms of decoding complexity and code overhead and compares measurements results obtained from our implementation with the analytical expressions.
Readers primarily interested in theoretical results and familiar with \ac{FEC} and \ac{RLNC} could therefore skip \secref{related} as well as some parts of \secref{code}.


\section{\ac{RLNC} and Related Work} \label{sec:related}

When data is distributed from one or more sources to one or more sinks using \ac{RLNC}, then it is \textit{encoded} at the sources to produce coded symbols and coding vectors that describe the encoding procedure. Together a coded symbol and a coding vector form a coded packet. When a sink has received \textit{enough} coded packets, it can \textit{decode} the original data. Additionally, received symbols can be recombined and thus \textit{recoded}, at any relaying nodes in between the sources and sinks.

\begin{figure}[htp]
\centering
$\underbrace{
\begin{array}{ |c|c|c| }
\hline
\mathrm{ \ Existing \ header \ } & \mathrm{\ Coding \ vector \ } & \mathrm{ \ Coded \ symbol \ }\\ \hline
\end{array}
}_{\mathrm{Coded \ packet}}$
\end{figure}

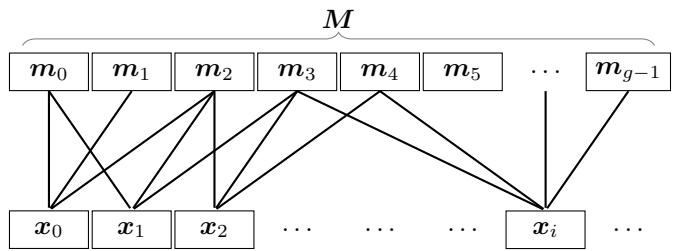
\begin{figure}[htp]
\centering

  \begin{tikzpicture}[scale=\scaletikz, node distance=1.1cm,auto ]

    \tikzstyle{invisible}=[minimum height=.5cm, minimum width=1.05cm]
    \tikzstyle{symbol}=[invisible, rectangle, draw=black]
    \tikzstyle{arrow}=[-,>=stealth,thick,shorten >=1pt,rectangle,thick,minimum size=4mm]

    \node[symbol] (s0) [] {$\vect{m}_0$};
    \node[symbol] (s1) [right of=s0] {$\vect{m}_1$};
    \node[symbol] (s2) [right of=s1] {$\vect{m}_2$};
    \node[symbol] (s3) [right of=s2] {$\vect{m}_3$};
    \node[symbol] (s4) [right of=s3] {$\vect{m}_4$};
    \node[symbol] (s5) [right of=s4] {$\vect{m}_5$};
    \node[invisible] (s6) [right of=s5] {$\hdots$};
    \node[symbol] (sg1) [right of=s6] {$\vect{m}_{g-1}$};

    \node[symbol,yshift=-1cm] (c0) [below of=s0] {$\vect{x}_0$};
    \node[symbol] (c1) [right of=c0] {$\vect{x}_1$};
    \node[symbol] (c2) [right of=c1] {$\vect{x}_2$};
    \node[invisible] (c3) [right of=c2] {$\hdots$};
    \node[invisible] (c4) [right of=c3] {$\hdots$};
    \node[invisible] (c5) [right of=c4] {$\hdots$};
    \node[symbol] (c6) [right of=c5] {$\vect{x}_i$};
    \node[invisible] (c7) [right of=c6] {$\hdots$};

    \path
    (s0.south) edge [arrow] node[] {} (c0.north)
    (s1.south) edge [arrow] node[] {} (c0.north)
    (s2.south) edge [arrow] node[] {} (c0.north);

    \path
    (s0.south) edge [arrow] node[] {} (c1.north)
    (s2.south) edge [arrow] node[] {} (c1.north)
    (s3.south) edge [arrow] node[] {} (c1.north);

    \path
    (s2.south) edge [arrow] node[] {} (c2.north)
    (s3.south) edge [arrow] node[] {} (c2.north)
    (s4.south) edge [arrow] node[] {} (c2.north);

    \path
    (s3.south) edge [arrow] node[] {} (c6.north)
    (s4.south) edge [arrow] node[] {} (c6.north)
    (s6.south) edge [arrow] node[] {} (c6.north)
    (sg1.south) edge [arrow] node[] {} (c6.north);

    \draw [gray,decorate,decoration={brace,amplitude=5pt}]
    (-.25,0.25)  -- (5.75,0.25)
    node [black,midway,above=3pt,xshift=0pt] {$\mat{M}$};
  \end{tikzpicture}
\caption{Coded symbols are created from the original data.}
\label{fig:coded_symbols}

\end{figure}

For practical reasons, the original data is typically divided into \textit{generations}~\cite{Chou03practicalnetwork} of size $g$. We denote the data in such a generation as $\mat{M}$. This ensures that coding can be performed over data of any size, and that the performance of \ac{RLNC} is independent of the data size. Each generation is divided into symbols, denoted as $\vect{m}_i$, and these symbols are then combined at random to create a coded symbol, denoted as $\vect{x}_i$. As all operations are performed over a \ac{FF} $\mathbb{F}_q$, the code is linear, and thus new valid coded symbols can be created from coded and non-coded symbols. \figref{coded_symbols} illustrates how the original symbols can be combined at random and provide an endless stream of coded symbols. The original data can be decoded by inverting the coding operations performed on the coded symbols. See~\cite{Neubauer:2007,primer} for introductions to \ac{FF} and \ac{RLNC}.

Dividing the data into generations reduces both the computational work and the decoding delay. Unfortunately, it also introduces the need for additional signaling~\cite{chunked_codes,yao}, as each of the generations must be decoded successfully before the original data is recovered fully. It also increases the probability that the sink receives linearly dependent symbols which adds to the overhead of the code.
This overhead is well understood for network typologies where (it can be assumed that) symbols are only received from sources that hold the original information~\cite{5634159,6188495,ICC2009}.
In such systems, the parameters of the code can be chosen so that the overhead tends to zero and can be ignored.
%
However, these parameters present a trade-off where higher values will generally result in lower code overhead but lower coding throughput~\cite{NCRealWorld}.
The coding throughput also depends on less deterministic parameters, e.g. the hardware platform, programming language, and implementation optimizations~\cite{shojania-parrallel,jcn2008,chinese-gfx,shojania-iphone,ICC2009}. Therefore a universally optimal set of values cannot be identified, as they depend on the system and on the target platform.

Some simplifications that can increase the coding throughput of \ac{RLNC} are binary, systematic, and sparse variants~\cite{binary,ICC2009,sparse,ICC2011}.
Binary codes are in widespread use and can obtain a low code overhead. They can be fast as operations in the binary field can be performed in parallel by all modern computers.
Using a systematic code comes with no cost in term of overhead, and can potentially provide a high gain in both encoding and decoding throughput. Unfortunately, it is not possible to us this approach at every node, but only at the sources. Thus there is no or little gain if recoding is performed, which is the main reason to use \ac{RLNC} in the first place.
Using a sparse random code provides similar benefits and drawbacks as a systematic code. It becomes impractical to perform recoding, and the gain in decoding throughput can be small or non-existing~\cite{NCpro2011A}.

Alternatively, the underlying code can be fundamentally modified or replaced to ensure a lower decoding complexity.
A noteworthy suggestion is to use a convolutional code as the underlying code~\cite{citeulike:10205307,citeulike:10205303} as they have been used in communication systems for many years. These efforts are still primarily theoretical as to the best of our knowledge currently no implementation of convolutional codes for \ac{NC} exists.
The work on perpetual codes~\cite{perpetual} is related to this work, since it uses a related fundamental concept, combined with a concatenated approach similarly to Raptor codes~\cite{raptor}. However the authors aim was to propose a cache-friendly rateless erasure code, and they did not consider recoding, which is a necessary feature when used in a system that exploits \ac{NC}.
%
We note that linear block codes and convolutional codes may in some cases be \textit{equivalent}, as they can describe codes with similar realizations using different terminology~\cite{Fragouli04aconnection,equivalent}.


%

Another direction in the search for improved trade-off between computational work and code overhead was suggested in~\cite{silva}.
Here the authors considered coding over several generations, called a \textit{random annex} code~\cite{DBLP:journals/corr/abs-1011-3498,yao}. Each generation is extended to include symbols from other generations and thus when a generation is decoded these extra symbols are released. This reduces the problem of ensuring that all generations are decoded, and thus the overhead. At the same time, it is less computationally demanding as the decoding is performed in an inner and an outer step. The approach is very useful for file transfers, but less so for streaming as the final decoding delay is high as generations are not decoded sequentially. Additionally, the problem of how recoding could be performed has so far not been considered.
Note that the idea of a \textit{random annex} can be applied to many underlying codes, including the perpetual code considered in this work.

\section{Code Operation} \label{sec:code}

This section introduces the code and the three operations, \textit{encoding}, \textit{decoding} and \textit{recoding}, that can be performed at nodes in the network. The notation used for analysis and algorithms is listed in Table~\ref{tab:notation}.
In vectors and matrices, we denote the first element with index zero. In some algorithms, the value $-1$ is used to denote non-valid or non-existing.

\begin{table}[htp]
\caption{Notation used for analysis and algorithms}
\label{tab:notation}
\centering
\begin{tabular}{l p{7cm}}
\toprule
Symbol & Definition  \\
\midrule
$g$ & Generation size \\
$q$ & Field size \\
$w$ & Coding vector width \\
$\mathbb{F}_q$ & A finite field with $q$ elements\\
$\vect{g}$ & Coding vector with $g$ elements, starting at element 0 \\
$\mat{G}$ & Matrix containing all received coding vectors \\
$\vect{x}$ & Coded symbol \\
$\mat{X}$ & Matrix containing all received symbol \\
$\vect{h}$ & Local recoding vector \\
\midrule
$\mat{G}_{i}$ & The $i$th row of the coding matrix\\
$\mat{G}_{i,j}$ & The index in row $i$ and column $j$ of the coding matrix.\\
$\mat{X}_{i}$ & The $i$th row of the symbol matrix\\
$\mat{X}_{i,j}$ & The index in row $i$ and column $j$ of the symbol matrix.\\
%
%
$p$ & Local variables for pivot indices \\

$?$ & A randomly drawn integer.\\
\bottomrule
\end{tabular}
\label{tab:algorithm_notation}
\end{table}

In \ac{RLNC}, the elements in the coding vector $\vect{g}$ are drawn completely at random, and thus each coded symbol is a combination of all the original symbols in one generation. This is not the case for the perpetual approach that we consider in this work. Instead, an element with index $p$ is chosen as the pivot and the following $w$ elements are drawn at random from $\field{F}_q$. We denote $w$ as the width of the coding vector. See~\figref{coding_matrix} for a small example of some resulting coding vectors.
%
%

\def\matrixCells{8}
\def\stepSize{.5}
\def\width{3}
\begin{figure}[htp]

\centering

  \begin{tikzpicture}[scale=\scaletikz]
    \draw[step=.5] (0,0) grid (4,4);
    \foreach \x/\y in {1/8, 2/7, 3/6, 4/5, 5/4, 6/3, 7/2, 8/1}
             {
               \node at (\x * \stepSize -.25 ,\y * \stepSize - .25) {1};
             }

             \foreach \x/\y in
                      {
                        2/8, 3/8, 4/8,
                        3/7, 4/7, 5/7,
                        4/6, 5/6, 6/6,
                        5/5, 6/5, 7/5,
                        6/4, 7/4, 8/4,
                        7/3, 8/3, 1/3,
                        8/2, 1/2, 2/2,
                        1/1, 2/1, 3/1
                      }
                      {
                        \pgfmathtruncatemacro{\xp}{\x-1}
                        \pgfmathtruncatemacro{\yp}{8-\y}
                        \node at (\x * \stepSize -.25 ,\y * \stepSize - .25) {$\gamma _{\yp,\xp}$};
                      }

                      \draw [gray,decorate,decoration={brace,amplitude=5pt}]
                      (.5,4.1)  -- (2.0,4.1)
                      node [black,midway,above=3pt,xshift=0pt] {$w$};

                      \draw [gray,decorate,decoration={brace,amplitude=5pt}]
                      (-.1,0)  -- (-.1,1.5)
                      node [black,midway,left=0pt, xshift=-3pt] {$w$};

                      \draw [gray,decorate,decoration={brace,amplitude=5pt}]
                      (4.0,-.1)  -- (0,-.1)
                      node [black,midway,below=3pt, xshift=0] {$g$};

                      \draw [gray,decorate,decoration={brace,amplitude=5pt}]
                      (4.1,4.0)  -- (4.1,0)
                      node [black,midway,right=3pt, xshift=0] {$g$};

  \end{tikzpicture}

\caption{All possible coding vectors, when $g=8$ and $w=3$. The $\gamma$'s denote randomly drawn elements from $\field{F}_q$.}
\label{fig:coding_matrix}

\end{figure}
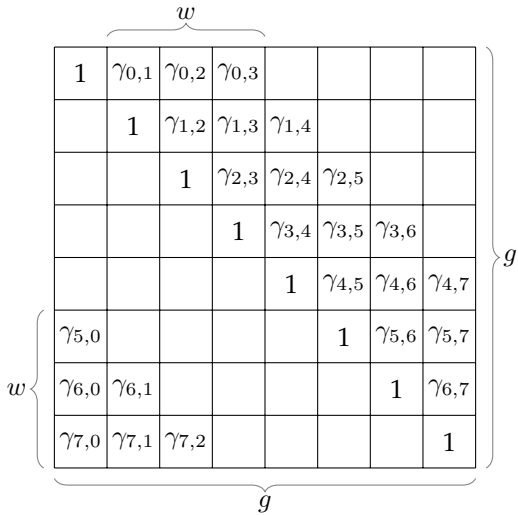

\subsection{Encoding} \label{subsec:encoding}

The data to be transmitted from the source is divided into generations, we denote the data in such a generation $\mat{M}$. Each generation is divided into $g$ symbols that are represented with one or more \ac{FF} elements in $\mathbb{F}_q$. The symbols are combined as specified by the coding vector $\vect{g}$ in order to create coded symbols $\vect{x}$.
\begin{align}
\vect{x} = \mat{M} \cdot \vect{g}
\label{eq:encode}
\end{align}

The construction of a coding vector $\vect{g}$ and the corresponding coded symbol $\vect{x}$ is described by \algref{alg:encode}.





\begin{algorithm}[htp]
  \caption{encode}
  \label{alg:encode}
  \KwIn{$\mat{M}$}
  $\vect{g} \leftarrow \vect{0}$\\
  $p \leftarrow (? \mod g)$\\
  $\vect{g}_p \leftarrow 1$\\
  \For{$i \in (p,p+w]$}
  {
    $\vect{g}_{(i \mod g)} \leftarrow (? \mod q)$
  }
  $\vect{x} \leftarrow \mat{M} \cdot \vect{g}$

  \Return $\vect{g}, \vect{x}$

\end{algorithm}

An index in the generation is drawn at random and used as the pivot, $p \in [0,g)$. The index in $\vect{g}$ that corresponds to this pivot element is set to one. For the subsequent $w$ indices in $\vect{g}$, an element is drawn at random from $\mathbb{F}_q$. The remaining elements in $\vect{g}$ are zeros. The resulting coding vector is of the form illustrated in \figref{coding_matrix}.
 To create a coded symbol, the coding vector is multiplied with the data, $\vect{x} = \mat{M} \cdot \vect{g}$. Together, the coding vector  $\vect{g}$ and coded symbol $\vect{x}$ form a coded packet.

It is trivial to represent the coding vector in a very compact way. Each coding vector can be represented by an index and $w$ scalars.
\begin{figure}[ht]
\centering
$ \begin{array}{| *{5}{c|} }
\hline
p & s_1 & s_2 & \hdots & s_w
\\ \hline
\end{array} $
\end{figure}
 The necessary bits for their representation is given by \eqref{vector_bits}. The index can take $g$ values and each of the $w$ elements can take $q$ values.
\begin{align}
|\vect{g}| = \log _2 (g) + w \cdot \log _2 (q) \label{eq:vector_bits} \ [\mathrm{bits}]
\end{align}

Coding vectors can be generated in slightly different ways depending on how $p$ is drawn and the size of $w$, see \tabref{modes}. The \textit{systematic} mode does not produce coding vectors of the specified form, but we include it for completeness.

\begin{table}[htp]
\centering
\setlength{\tabcolsep}{3pt}
\caption{Different encoding modes.}
\label{tab:modes}
\begin{tabular}{l p{5cm} l }
\toprule
Mode & $p$ drawn & $w$ \\
\midrule
Random & random  $ \in [0,g)$ & $0<w<g$ \\
Sequential & sequentially looping from 0 to g-1 & $0<w<g$ \\
Systematic & sequentially from 0 to g-1, subsequently drawn at random $ \in [0,g)$ & $w=0$ \\
\bottomrule
\end{tabular}
\end{table}


\subsection{Decoding} \label{sec:decoding}

A node that receives coded packets can decode the original data by collecting the coded symbols in $\mathat{X}$ and the coding vectors in $\mathat{G}$. The original information, $\mat{M}$ can be found as in \eqref{decode}, provided that $\mathat{G}$ is invertible and thus has full rank.
\begin{align}
\mat{M} = \mathat{X} \cdot \mathat{G}^{-1}
\label{eq:decode}
\end{align}

To decode the original data in $\mathat{M}$, $\mathat{G}$ must be reduced to identity form by performing basic row operations that are simultaneously performed on $\mathat{M}$.
When it is not possible to fully decode a symbol upon reception, then it is partially decoded, and stored for later processing. This is referred to as \textit{on-the-fly} decoding. When enough symbols have been received so that $\mathat{G}$ has full rank, all received symbols can be fully decoded and the original data can be retrieved, we refer to this as \textit{final} decoding.

Unlike \ac{RLNC} and \ac{SRLNC}, this perpetual approach defines the location of the non-zero values in the coding vector. This makes it possible to decode symbols efficiently and without the problematic fill-in that can be observed during decoding and recoding \ac{SRLNC}~\cite{NCpro2011A}.

\subsubsection{On-the-fly Decoding}

When a new coded packet arrives, its coding vector is inserted into the decoding matrix iff. it has a \textit{pivot candidate} that was not previously identified. We distinguish between \textit{pivot} and \textit{pivot candidate} as the element that is used as the pivot may only be found during the final decoding.
Otherwise the previously received symbol with the same pivot candidate is subtracted from the new symbol, and the pivot candidate of the new symbol is changed. This is repeated until a new pivot candidate is identified. If the coding vector is reduced to the zero vector, the symbol is discarded.

\begin{figure*}[htp]
  \centering
             \begin{tikzpicture}[scale=\scaletikz]
               \draw[step=.5] (0,0) grid (4,4);
               \foreach \x/\y/\char in
                        {
                          1/8/1, 2/8/$\gamma_{\yp,\xp}$, 3/8/$\gamma_{\yp,\xp}$, 4/8/$\gamma_{\yp,\xp}$,
                          2/7/1, 3/7/$\gamma_{\yp,\xp}$, 4/7/$\gamma_{\yp,\xp}$, 5/7/$\gamma_{\yp,\xp}$,
                          8/1/1, 1/1/$\gamma_{\yp,\xp}$, 2/1/$\gamma_{\yp,\xp}$, 3/1/$\gamma_{\yp,\xp}$
                        }
                        {
                          \pgfmathtruncatemacro{\xp}{\x-1}
                          \pgfmathtruncatemacro{\yp}{8-\y}
                          \node at (\x * \stepSize -.25 ,\y * \stepSize - .25) {\char};
                        }

               \foreach \x/\y/\char in
                        {
                          -9/8/1, -8/8/$\hat{\gamma}_{\xp}$, -7/8/$\hat{\gamma}_{\xp}$, -6/8/$\hat{\gamma}_{\xp}$,
                          -8/7/$\hat{\gamma}_{\xp}$, -7/7/$\hat{\gamma}_{\xp}$, -6/7/$\hat{\gamma}_{\xp}$,
                          -7/6/$\hat{\gamma}_{\xp}$, -6/6/$\hat{\gamma}_{\xp}$, -5/6/$\hat{\gamma}_{\xp}$
                        }
                        {
                          \pgfmathtruncatemacro{\xp}{\x+9}
                          \pgfmathtruncatemacro{\yp}{8-\y}
                          \node at (\x * \stepSize -.25 ,\y * \stepSize - .25) {\char};
                        }

               \foreach \x/\y/\char in
                        {
                          3/6/1, 4/6/$\gamma_{\yp,\xp}$, 5/6/$\gamma_{\yp,\xp}$
                        }
                        {
                          \definecolor{myorange}{rgb}{.8886, 0.5782, 0.2645}
                          \draw[fill=myorange] (\x * \stepSize - .5, \y * \stepSize) --
                          (\x * \stepSize , \y * \stepSize ) --
                          (\x * \stepSize , \y * \stepSize -.5) --
                          (\x * \stepSize -.5, \y * \stepSize -.5) -- cycle;

                          \pgfmathtruncatemacro{\xp}{\x-1}
                          \pgfmathtruncatemacro{\yp}{8-\y}
                          \node [] at ( \x * \stepSize -.25 ,\y * \stepSize - .25) {\char};
                        }

               \draw[step=.5] (-5,2.49) grid (-1,4);

               \foreach \x/\y in {-9/7, -8/6 }
                        {
                          \definecolor{mygreen}{rgb}{0.7098, 0.7961, 0.5916}
                          \draw[fill=mygreen] (\x * \stepSize - .5, \y * \stepSize) --
                          (\x * \stepSize , \y * \stepSize ) --
                          (\x * \stepSize , \y * \stepSize -.5) --
                          (\x * \stepSize -.5, \y * \stepSize -.5) -- cycle;

                          \node at (\x * \stepSize -.25 ,\y * \stepSize - .25) {0};
                        }

                      \draw [->] (-.25,3.75)  -- node[above]{2.} (-.75,3.75) ;
                      \draw [->] (-.25,3.25)  -- node[above]{4.} (-.75,3.25) ;
                      \draw [<-] (-.25,2.75)  -- node[above]{6.} (-.75,2.75) ;

                      \draw [*->] (-5.75,3.8) -- node[above]{1.} (-5.25,3.8) ;
                      \draw [->]  (-5.25,3.7) arc (-90:90:-6pt) node[above, xshift=-.5cm]{3.} ;
                      \draw [->] (-5.25,3.2) arc (-90:90:-6pt) node[above, xshift=-.5cm]{5.} ;

             \end{tikzpicture}

\caption{\textit{On-the-fly} decoding of a received coded packet. The right hand-side matrix is the decoding matrix $\mathat{G}$. The left hand-side matrix shows the coding vector of the incoming symbol as it is decoded. The $\gamma$'s denotes random field elements. The filled circle and arrow indicate the original coding vector of the incoming packet. The straight lines indicate which rows are substituted into the coding vector. The arcs indicate the decoding steps.}
\label{fig:decode1}
\end{figure*}
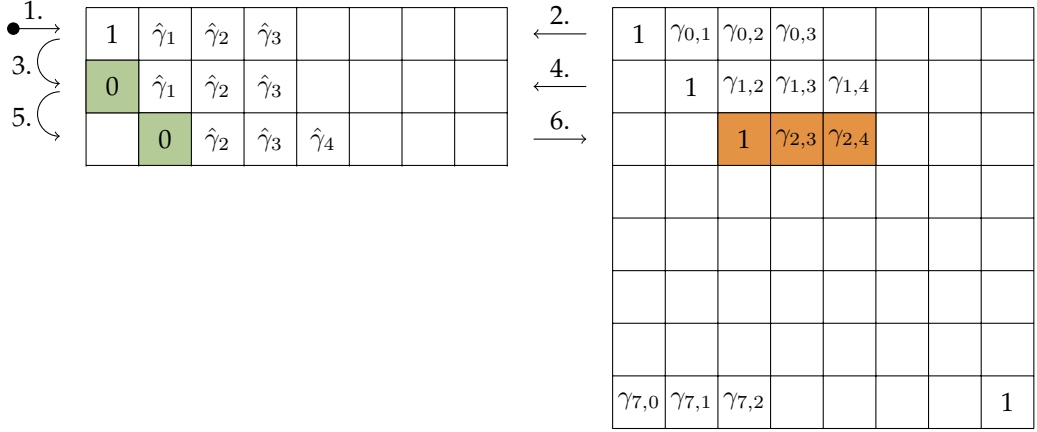

In \figref{decode1}, we assume that three coded packets have been received and their coding vectors have been inserted into the decoding matrix. The pivot candidates of the received packets are zero, one, and seven, respectively. Subsequently, a coded packet with pivot candidate zero is received. This is denoted with a filled circle and arrow pointing to the coding vector of the packet in the left hand-side matrix. A symbol with the same pivot candidate have already been identified. Therefore, the existing row zero is subtracted from the incoming packet. This is denoted with the arrow pointing left into the left hand-side matrix. The element that initially was the pivot candidate is now zero and an element to the right has now become the pivot candidate. This step is repeated for the new pivot candidate and row one is subtracted from the incoming packet and element two becomes the pivot candidate. As this pivot candidate was previously not identified, the coding vector is inserted into the decoding matrix, which is marked with orange and the arrow pointing right into the decoding matrix.

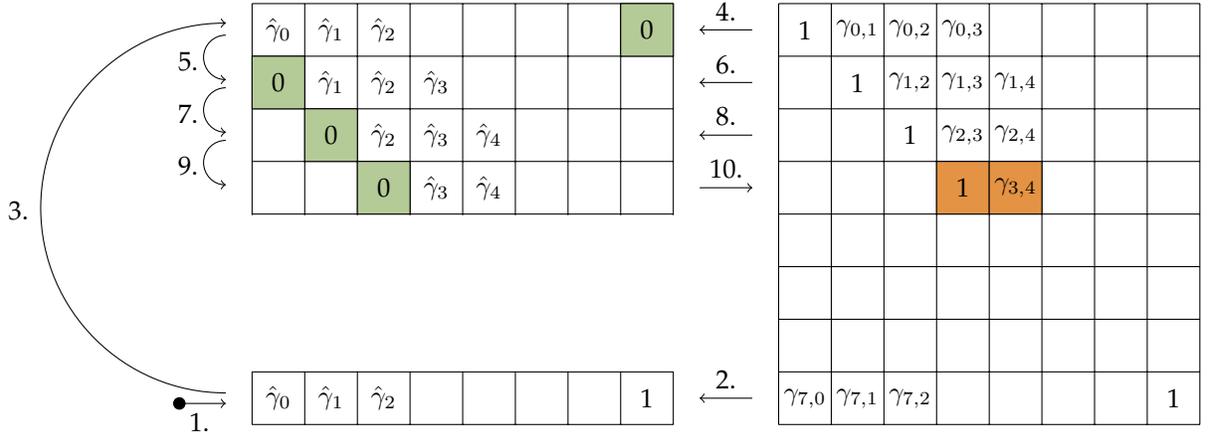
\begin{figure*}[htp]
  \centering
             \begin{tikzpicture}[scale=\scaletikz]

               \draw[step=.5] (-5,1.99) grid (-1,4);
               \draw[step=.5] (-5,0) grid (-1,.5);

               \draw[step=.5] (0,0) grid (4,4);
               \foreach \x/\y/\char in
                        {
                          1/8/1, 2/8/$\gamma_{0,1}$, 3/8/$\gamma_{0,2}$, 4/8/$\gamma_{0,3}$,
                          2/7/1, 3/7/$\gamma_{1,2}$, 4/7/$\gamma_{1,3}$, 5/7/$\gamma_{1,4}$,
                          3/6/1, 4/6/$\gamma_{2,3}$, 5/6/$\gamma_{2,4}$,
                          8/1/1, 1/1/$\gamma_{7,0}$, 2/1/$\gamma_{7,1}$, 3/1/$\gamma_{7,2}$,
                          -9/1/$\hat{\gamma}_0$, -8/1/$\hat{\gamma}_1$, -7/1/$\hat{\gamma}_2$, -2/1/1,
                          -9/8/$\hat{\gamma}_0$, -8/8/$\hat{\gamma}_1$, -7/8/$\hat{\gamma}_2$,
                          -8/7/$\hat{\gamma}_1$, -7/7/$\hat{\gamma}_2$, -6/7/$\hat{\gamma}_3$,
                          -7/6/$\hat{\gamma}_2$, -6/6/$\hat{\gamma}_3$, -5/6/$\hat{\gamma}_4$,
                          -6/5/$\hat{\gamma}_3$, -5/5/$\hat{\gamma}_4$
                        }
                        {
                          \node at (\x * \stepSize -.25 ,\y * \stepSize - .25) {\char};
                        }

               \foreach \x/\y/\char in
                        {
                          4/5/1, 5/5/$\gamma_{3,4}$
                        }
                        {
                          \definecolor{myorange}{rgb}{.8886, 0.5782, 0.2645}
                          \draw[fill=myorange] (\x * \stepSize - .5, \y * \stepSize) --
                          (\x * \stepSize , \y * \stepSize ) --
                          (\x * \stepSize , \y * \stepSize -.5) --
                          (\x * \stepSize -.5, \y * \stepSize -.5) -- cycle;

                          \node [] at (\x * \stepSize -.25 ,\y * \stepSize - .25) {\char};
                        }

               \foreach \x/\y in {-2/8, -9/7, -8/6, -7/5}
                        {
                          \definecolor{mygreen}{rgb}{0.7098, 0.7961, 0.5916}
                          \draw[fill=mygreen] (\x * \stepSize - .5, \y * \stepSize) --
                          (\x * \stepSize , \y * \stepSize ) --
                          (\x * \stepSize , \y * \stepSize -.5) --
                          (\x * \stepSize -.5, \y * \stepSize -.5) -- cycle;

                          \node [] at (\x * \stepSize -.25 ,\y * \stepSize - .25) {0};
                        }

                      \draw [->] (-.25,0.25)  -- node[above]{2.} (-.75,0.25) ;
                      \draw [->] (-.25,3.75)  -- node[above]{4.} (-.75,3.75) ;
                      \draw [->] (-.25,3.25)  -- node[above]{6.} (-.75,3.25) ;
                      \draw [->] (-.25,2.75)  -- node[above]{8.} (-.75,2.75) ;
                      \draw [<-] (-.25,2.25)  -- node[above]{10.} (-.75,2.25) ;

                      \draw [*->] (-5.75,.2)  -- node[below]{1.} (-5.25,.2);
                      \draw [->]  (-5.25,.3) arc (-90:-270:50pt) node[left, xshift=-2.5cm, yshift=-2.5cm]{3.};
                      \draw [->]  (-5.25,3.7) arc (-90:90:-6pt) node[above, xshift=-.5cm]{5.};
                      \draw [->]  (-5.25,3.2) arc (-90:90:-6pt) node[above, xshift=-.5cm]{7.};
                      \draw [->]  (-5.25,2.7) arc (-90:90:-6pt) node[above, xshift=-.5cm]{9.};

             \end{tikzpicture}
\caption{\textit{On-the-fly} decoding similar to \figref{decode1}, but the pivot candidate wraps around the end of the decoding matrix.}
\label{fig:decode2}

\end{figure*}

A special case is when the on-the-fly phase causes the pivot candidate to wrap around to the start of the coding vector.
An example of this is illustrated in \figref{decode2}. The incoming packet has pivot candidate seven for which a pivot candidate has already been identified in $\mathat{G}$. Thus row seven in  $\mathat{G}$ is subtracted from the incoming packet.
If the last element in the coding vector is reduced to the zero vector, the first element in the vector is considered next and becomes the pivot candidate.
In this case, the resulting coding vector has a zero at index seven and thus the pivot candidate is now index zero. The packet is then further reduced similarly to the example in \figref{decode1}.

\begin{algorithm}[htp]
  \caption{forwardSubstitute}
  \label{alg:forwardSubstitute}
  \KwIn{$\vect{g}, \vect{x}$}

  \While{$\vect{g} \neq \vect{0}$}
      {
        $p \leftarrow \mathrm{pivot}(\vect{g})$

        \If{$\mat{G}_p \neq \vect{0} $}
           {
             $\vect{g} \leftarrow \vect{g} \cdot \frac{1}{\vect{g}_p}  \oplus \mat{G}_{p}$ \\
             $\vect{x} \leftarrow \vect{x}  \cdot \frac{1}{\vect{g}_p} \oplus \mat{X}_{p}$ \Comment substitute into new packet
           }
           \Else
               {
                 $\mat{G}_p \leftarrow \vect{g} \cdot \frac{1}{\vect{g}_p} $ \\
                 $\mat{X}_p \leftarrow \vect{X} \cdot \frac{1}{\vect{g}_p}$ \Comment insert new packet\\
                 {\bf return} $p$
               }
      }
      {\bf return} $-1$
\end{algorithm}


In \algref{alg:forwardSubstitute}, the existing row with the same pivot candidate is substituted into the received symbol, unless the received coding vector has been reduced to the zero vector. If a new pivot candidate is identified, then the coding vector and symbol are inserted into the respective matrices. Importantly, this algorithm guarantees that $w$ will not increase during decoding.

The coding vector can be reduced to the zero vector if it is a linear combination of previously received coding vectors. It is possible to end in a dead-lock where a sequence of rows is repeatedly subtracted from the new packet. To avoid this, decoding should be terminated after some attempts and the packet discarded. From practical experiments, it has been determined that decoding can be terminated after $2g$ or $3g$ iterations. To avoid wasting operations on such cases, row operations can first be performed on the coding vector and then repeated on the coded symbol~\cite{NCpro2011A}.
In both cases, the overhead arises because the symbol is a linear combination of already received symbols.

A simple optimization in cases where the $w$ of the incoming packet is lower than the $w$ of the held symbol with the same pivot candidate, is to simply swap these two to guarantee that $w$ is never increased. Our current implementation does not support this and we leave it to future work to test whether this increases the decoding throughput. However, previous experiments showed that such optimizations can introduce a high cost in terms of bookkeeping~\cite{NCpro2011A}.

\subsubsection{Final Decoding}

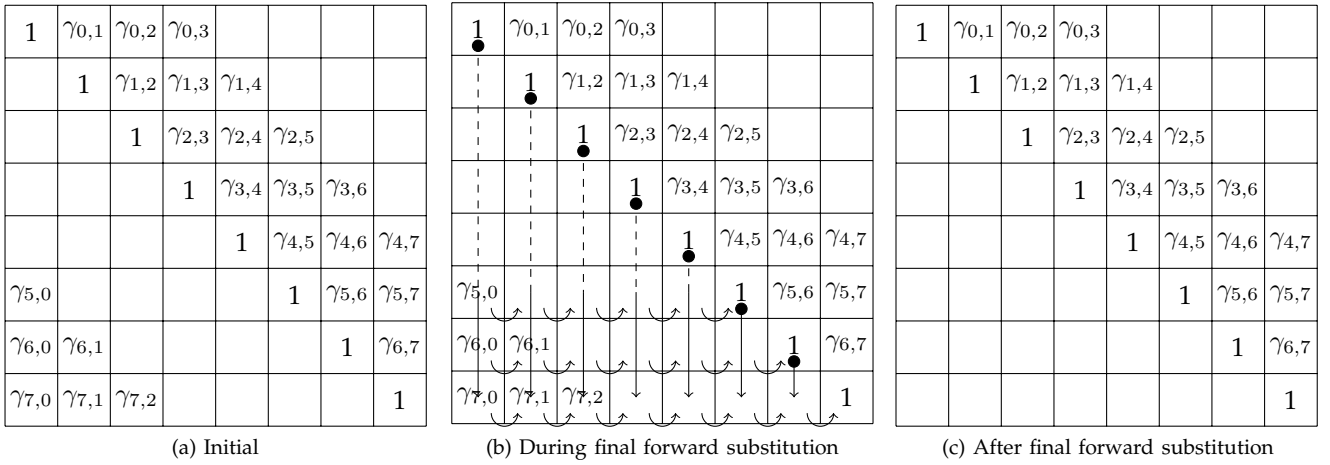
\begin{figure*}[htp]
 \centering
  \subfloat[Initial]
           {
             \label{fig:final_initial}
             \begin{tikzpicture}[scale=\scaletikz]
               \draw[step=.5] (0,0) grid (4,4);
               \foreach \x/\y in {1/8, 2/7, 3/6, 4/5, 5/4, 6/3, 7/2, 8/1}
                        {
                          \node at (\x * \stepSize -.25 ,\y * \stepSize - .25) {1};
                        }

               \foreach \x/\y\char in
                        {
                          2/8/$\gamma_{0,1}$, 3/8/$\gamma_{0,2}$, 4/8/$\gamma_{0,3}$,
                          3/7/$\gamma_{1,2}$, 4/7/$\gamma_{1,3}$, 5/7/$\gamma_{1,4}$,
                          4/6/$\gamma_{2,3}$, 5/6/$\gamma_{2,4}$, 6/6/$\gamma_{2,5}$,
                          5/5/$\gamma_{3,4}$, 6/5/$\gamma_{3,5}$, 7/5/$\gamma_{3,6}$,
                          6/4/$\gamma_{4,5}$, 7/4/$\gamma_{4,6}$, 8/4/$\gamma_{4,7}$,
                          7/3/$\gamma_{5,6}$, 8/3/$\gamma_{5,7}$, 1/3/$\gamma_{5,0}$,
                          8/2/$\gamma_{6,7}$, 1/2/$\gamma_{6,0}$, 2/2/$\gamma_{6,1}$,
                          1/1/$\gamma_{7,0}$, 2/1/$\gamma_{7,1}$, 3/1/$\gamma_{7,2}$
                        }
                        {
                          \node at (\x * \stepSize -.25 ,\y * \stepSize - .25) {\char};
                        }
             \end{tikzpicture}
           }
  \subfloat[During final forward substitution]
           {
             \label{fig:final_forward}
             \begin{tikzpicture}[scale=\scaletikz]
               \draw[step=.5] (0,0) grid (4,4);
               \foreach \x/\y in {1/8, 2/7, 3/6, 4/5, 5/4, 6/3, 7/2, 8/1}
                        {
                          \node at (\x * \stepSize -.25 ,\y * \stepSize - .25) {1};
                        }

               \foreach \x/\y\char in
                        {
                          2/8/$\gamma_{0,1}$, 3/8/$\gamma_{0,2}$, 4/8/$\gamma_{0,3}$,
                          3/7/$\gamma_{1,2}$, 4/7/$\gamma_{1,3}$, 5/7/$\gamma_{1,4}$,
                          4/6/$\gamma_{2,3}$, 5/6/$\gamma_{2,4}$, 6/6/$\gamma_{2,5}$,
                          5/5/$\gamma_{3,4}$, 6/5/$\gamma_{3,5}$, 7/5/$\gamma_{3,6}$,
                          6/4/$\gamma_{4,5}$, 7/4/$\gamma_{4,6}$, 8/4/$\gamma_{4,7}$,
                          7/3/$\gamma_{5,6}$, 8/3/$\gamma_{5,7}$, 1/3/$\gamma_{5,0}$,
                          8/2/$\gamma_{6,7}$, 1/2/$\gamma_{6,0}$, 2/2/$\gamma_{6,1}$,
                          1/1/$\gamma_{7,0}$, 2/1/$\gamma_{7,1}$, 3/1/$\gamma_{7,2}$
                        }
                        {
                          \node at (\x * \stepSize -.25 ,\y * \stepSize - .25) {\char};
                        }

               \foreach \x/\y in
                        {
                          1/8, 2/7, 3/6, 4/5, 5/4
                        }
                        {
                          \draw [*-,dashed]  (\x * \stepSize -.25 ,\y * \stepSize - .35) -- (\x * \stepSize -.25 , 3* \stepSize - .25) ;
                        }

               \foreach \x/\y in
                        {
                          1/3, 2/3, 3/3, 4/3, 5/3
                        }
                        {
                          \draw [->]  (\x * \stepSize -.25 ,\y * \stepSize - .25) -- (\x * \stepSize -.25 , 1* \stepSize - .25) ;
                        }

               \foreach \x/\y in
                        {
                          6/3, 7/2
                        }
                        {
                          \draw [*->]  (\x * \stepSize -.25 ,\y * \stepSize - .35) -- (\x * \stepSize -.25 , 1* \stepSize - .25) ;
                        }

               \foreach \x/\y in
                        {
                          1/3, 2/3, 3/3, 4/3, 5/3,
                          1/2, 2/2, 3/2, 4/2, 5/2, 6/2,
                          1/1, 2/1, 3/1, 4/1, 5/1, 6/1, 7/1
                        }
                        {
                          \pgfmathsetmacro{\tmp}{.25*\stepSize}
                          \draw [->]  (\x * \stepSize -0.125 ,\y * \stepSize - .4) arc (-180:0:\tmp);
                        }

             \end{tikzpicture}
           }
%
%
%
   \subfloat[After final forward substitution]
            {
             \label{fig:final_forward_2}
              \begin{tikzpicture}[scale=\scaletikz]
                \draw[step=.5] (0,0) grid (4,4);
                \foreach \x/\y in {1/8, 2/7, 3/6, 4/5, 5/4, 6/3, 7/2, 8/1}
                         {
                           \node at (\x * \stepSize -.25 ,\y * \stepSize - .25) {1};
                         }

                \foreach \x/\y/\char in
                         {
                          2/8/$\gamma_{0,1}$, 3/8/$\gamma_{0,2}$, 4/8/$\gamma_{0,3}$,
                          3/7/$\gamma_{1,2}$, 4/7/$\gamma_{1,3}$, 5/7/$\gamma_{1,4}$,
                          4/6/$\gamma_{2,3}$, 5/6/$\gamma_{2,4}$, 6/6/$\gamma_{2,5}$,
                          5/5/$\gamma_{3,4}$, 6/5/$\gamma_{3,5}$, 7/5/$\gamma_{3,6}$,
                          6/4/$\gamma_{4,5}$, 7/4/$\gamma_{4,6}$, 8/4/$\gamma_{4,7}$,
                          7/3/$\gamma_{5,6}$, 8/3/$\gamma_{5,7}$,
                          8/2/$\gamma_{6,7}$
                         }
                         {
                           \node at (\x * \stepSize -.25 ,\y * \stepSize - .25) {\char};
                         }
              \end{tikzpicture}
            }
%
%

%
%
%
  \caption{The decoding matrix $\mathat{G}$ at various states of the final decoding. The dotted part of the arrows indicate rows where no substitution is needed. The arching arrows show how the pivot candidate moves towards the diagonal.}
  \label{fig:final_decoding}
\end{figure*}

When a pivot candidate has been identified for all rows, final decoding is performed by forward substitution and backwards substitution.
Initially, the decoding matrix has a form similar to that shown in \figref{final_initial}. Note that some of the elements $\gamma_{i,j}$ might be zero.
It should also be noted that even though a pivot candidate has been identified for all rows, this does not guarantee that the decoding matrix has full rank. Therefore it is important to perform the final decoding 
in a way that ensures that the decoding matrix is not left in a state where future decoding becomes impossible or problematic.

To bring the matrix onto echelon form, forward substitution is performed on the non-zero elements in the lower left corner of \figref{final_initial}. When forward substitution is performed on the first column, non-zero elements can be introduced in the lower $w$ rows and further substitution becomes necessary as illustrated on \figref{final_forward}. After the forward substitution step, the decoding matrix is brought onto echelon form in \figref{final_forward_2}.




The final forward step in \algref{alg:final_forward} ensures that the decoding matrix is always left in a valid state even if partial final decoding occurs. This happens in cases where it turns out that the decoding matrix does not have full rank, even though a pivot candidate for each row was identified.
\begin{algorithm}[htp]
  \caption{finalForward}
  \label{alg:final_forward}
  \For{$i \in  [0, \hdots, g )$}
    {
      \For{$j \in  [i, \hdots, g )$}
        {
          \If{$\mat{G}_{j,i} \neq 0$}
             {
               \If{$i = j$}
                  {
                    $\mat{G}_{i} \leftarrow \mat{G}_i \cdot \frac{1}{\mat{G}_{i,i}}$ \\
                    $\mat{X}_{i} \leftarrow \mat{X}_i \cdot \frac{1}{\mat{G}_{i,i}}$  \Comment normalize
                  }
               \If{$i \neq j$}
                  {
                    $\mat{G}_{i} \leftrightarrow \mat{G}_j \cdot \frac{1}{\mat{G}_{j,i}}$ \\
                    $\mat{X}_{i} \leftrightarrow \mat{X}_j \cdot \frac{1}{\mat{G}_{j,i}}$ \Comment normalize and swap
                  }
                  \For{$k \in [\max (j+1, g-w), g )$}
                    {
                      \If{$\mat{G}_{k,i} \neq 0$}
                         {
                           $\mat{G}_k \leftarrow \mat{G}_k \oplus \mat{G}_i$ \\
                           $\mat{X}_k \leftarrow \mat{X}_k \oplus \mat{X}_i$ \Comment substitute down
                         }
                    }
                    {\bf break} \Comment found a pivot, skip to the next
                  }
                  \Else
                      {
                        \If{$j = (g-1)$}
                           {
                             \If{$i \neq (g-1)$}
                                {
                                  \If{$\mat{G}_i \neq \mat{G}_{i+1}$}
                                     {

                                       $\mat{G}_{i+1} \leftarrow \mat{G}_{i+1} \oplus \mat{G}_{i}$\\
                                       $\mat{X}_{i+1} \leftarrow \mat{X}_{i+1} \oplus \mat{X}_{i}$\\
                                       \Comment add to below symbols
                                     }
                                }
                                $\mat{G}_{i} \leftarrow \vect{0}$ \\
                                $\mat{X}_{i} \leftarrow \vect{0}$ \Comment discard symbol
                           }
                      }
               }
        }
\end{algorithm}

In \algref{alg:final_forward}, a pivot element must be defined for each column ($i$). If no such pivot element can be found, then it means that none of the received symbols can be used to decode the corresponding row, and we need to receive additional symbols. Therefore we traverse all the rows ($j$) from the diagonal and down, as we know that for all rows above a pivot index have already been identified. When we find a row for which the current pivot element is non-zero, we swap it with the correct row if it is not already at the correct position. We then forward substitute into the rows below. If we iterate to the last row without identifying a pivot element, then we cannot decode the target row, and we discard the symbol that is incorrectly located on row $i$. However, we do not want to discard useful symbols, therefore we check if the coding vector for row $i$ is equal to the row below, if not we simply add it to the row below. If the two symbols were identical, the result would be the $\vect{0}$ vector and we would discard two rows instead of one. Finally, if we are looking for the pivot element for the final column, there are no rows below our target row, and we simply discard without checking.
In this way, the decoding matrix will always be brought as close to echelon form as possible.

If the rank of the matrix is full after the forward substitution, then standard backward substitution is performed to bring the decoding matrix to reduced echelon form and decode the original data.
\begin{algorithm}[htp]
  \caption{finalBackward}
  \label{alg:final_backward}

  \For{$i \in  (g, \hdots, 0) $}
  {
    \For{$j \in  (i, \hdots , \max (i-w,0) ] $}
  {
    \If{$\mat{G}_{j,i} \neq 0 $}
       {
         $\mat{G}_j \leftarrow \mat{G}_j \oplus \mat{G}_i \cdot \mat{G}_{j,i}$\\
         $\mat{X}_j \leftarrow \mat{X}_j \oplus \mat{X}_i \cdot \mat{G}_{j,i}$
       }
  }
  }
\end{algorithm}

Starting from the bottom, all rows are used to remove any remaining non-zeros in the rows above. Note that for each index, it is only necessary to inspect the above $w$ rows as all other rows are guaranteed to be zero due to the form of the decoding matrix.
Algorithms \ref{alg:forwardSubstitute}-\ref{alg:final_backward} can be combined to create the decoder in Algorithm~\ref{alg:decode}.
\begin{algorithm}[htp]
  \caption{decode}
  \label{alg:decode}
  \KwIn{$\vect{g}, \vect{x}$}


  \If{$\mathrm{forwardSubstitute}(\vect{g}, \vect{x}) \neq -1 $}
     {

  \If{$\mathrm{rank}(\mat{G}) = g$}
     {
       finalForward()
     }

  \If{$\mathrm{rank}(\mat{G}) = g$}
     {
       finalBackward()
     }

     }

  \Return $\mathrm{rank}(\mat{G})$

\end{algorithm}

 When a new packet arrives, it is first forward substituted. If a new pivot element is identified, the coding vector and the coded symbol are inserted into the decoding matrix. When the rank of the decoding matrix is full, final decoding is attempted using forward substitution. This might initially fail, but when it succeeds final backwards substitution is performed and the original data in the generation is decoded.

\subsection{Recoding}

When two or more coded or non-coded symbols have been received, they can be combined by recoding.
This can be described by \eqref{recode1} and (\ref{eq:recode2}) where the collected coding vectors and coded symbols are combined as defined by $\vect{h}$ of length $g'$, where $g'$ is the number of received symbols. Then $\vecttilde{x}$ and $\vecttilde{g}$ together form a recoded packet.
\begin{align}
\vecttilde{g} = \mathat{G} \cdot \vect{h} \label{eq:recode1}\\
\vecttilde{x} = \mathat{X} \cdot \vect{h} \label{eq:recode2}
\end{align}

 In classical \ac{RLNC}, coded packets are accumulated and recoding is performed as a separate operation which results in a significant computational load, we denote this type of recoding \textit{active} recoding.
As explained in \cite{ICC2011}, this form of recoding is not suitable when the code is sparse, because the recoded symbol will become denser with high probability~\cite{Ingram_minimumdegree,NCpro2011A}. To combat this problem, we introduce a new type of recoding called \textit{passive} recoding.

\subsubsection{Active Recoding}

%

Combining all collected packets completely at random, as in standard \ac{RLNC}, results in recoded packets where the non-zero elements are no longer confined to $w$ elements.
%
%
If we instead pick packets that have similar pivot elements, then in the worst case the resulting coded packet will only have slightly more non-zero elements $w'$ than that of the original coding vectors.
This decreases the freedom in recoding, but allows us to maintain the sparsity in recoded packets. Unfortunately, such an approach significantly increases the complexity of recoding as it introduces a search for an appropriate set of coding vectors. Additionally, it is more deterministic than the standard recoding approach, and thus great care must be taken to avoid generating more linearly dependent symbols.



\subsubsection{Passive Recoding}

When \textit{on-the-fly} decoding is performed, previously received symbols are subtracted from an incoming symbol to partially decode it. This combining of packets can also be considered as recoding and therefore the operations can be reused in order to reduce the computational load of recoding.

If the operations performed on the received symbols are tracked, a symbol where a sufficient number of operations have been performed can be used as a recoded symbol.
%
%
One way is to keep a list for each received symbol, to record what symbols are substituted into the symbol. However, this could become unfeasible if $g$ is high. It is simpler to hold an integer for each symbol that is used to count the number of other symbols that have been substituted into the symbol. It is important to remember that during decoding we attempt to decode the symbols, therefore symbols that have been reduced \textit{too much} should not be used as recoded symbols directly. We note that this \textit{passive} approach can also be used for other codes.



\subsubsection{Active plus Passive Recoding}

To combine the two types of recoding we can monitor the passive recoding. If some neighboring set of packets combined meet our criteria for row operations, we can combine these by \textit{actively} recoding them and thus obtain a recoded symbol. With this hybrid approach, we can recode symbols whenever we need them and still reduce the computation work associated with recoding.

\subsubsection{Re-encoding}

When a receiver has decoded a generation, it can encode packets the same way as the original source. This is sometimes referred to as recoding, which we believe is misleading, and instead denote this re-encoding to distinguish this from encoding at the original source.

\subsubsection{Implementation}

In our implementation we have chosen to implement a simple version of the \textit{active plus passive recoding}. The primary reason is to reuse the operations performed during decoding and at the same time allow recoding to be performed when and as much as desired. Another important consideration is to avoid introducing a deterministic behavior when recoding.

\begin{algorithm}[htp]
  \caption{recode}
  \label{alg:recode}
  \KwIn{$\mathat{G}, \mathat{X}$}

  \If{$\mathrm{rank}(\mathat{G}) = 0$}
     {
       \Return $-1$ \Comment no symbols available
     }

  $p \leftarrow (? \mod g)$\\
  \While{$\mathat{G}_{p,p} = 0 $}
        {
          $p \leftarrow (? \mod g)$ \Comment find pivot index
        }

  $\vect{h}_p \leftarrow 1$\\
  \For{$i \in (p,p+w]$}
  {
    \If{$\mathat{G}_{i,i} \neq 0 $}
       {
         $\vect{h}_{(i \mod g)} \leftarrow (? \mod q)$ \Comment draw $w$ elements
       }
  }
  $\vecttilde{g} \leftarrow \mathat{G} \cdot \vect{h}$\\
  $\vecttilde{x} \leftarrow \mathat{X} \cdot \vect{h}$ \Comment perform recoding

  \Return $\vect{g}, \vect{x}$

\end{algorithm}

First, row indices are drawn at random until a row that is non-zero is identified. Then for each of the following $w$ rows that are non-zero, a random coefficient is drawn which defines the recoding vector $\vect{h}$. The remaining indices in $\vect{h}$ are zeros. The new coding vector and coded symbol are then computed as  $\vecttilde{g} = \mathat{G} \cdot \vect{h}$ and $\vecttilde{x} = \mathat{X} \cdot \vect{h}$, respectively.

This approach ensure that the width of the recoded vector $w' \leq 2w$. However, it is worth observing that if a symbol with a higher $w$ is received, the size of $w$ can be reduced during the forward substitution. This would happen more frequently if the width of the received and the existing row is compared as mentioned in \secref{decoding}.

\section{Analysis and Experiments} \label{sec:analysis}

In this section we present analytical and experimental results on the code overhead, complexity and throughput. To verify the analytical expressions, we have implemented the proposed code in C++~\cite{kodo_perpetual}. This also provides us with the possibility to report on encoding and decoding throughput which is a more interesting parameter that defines the computational load at the coding nodes. The current implementation is well tested and we believe that it provides a good trade-off between simplicity and throughput. As the code is available under a research friendly license, we encourage suggestions that can improve the throughput or simplify the implementation.



\subsection{Overhead} \label{sec:code_overhead}



%
%
%
The code overhead depends on the field size, density, generation size and possibly other factors.
From standard \ac{RLNC}, we have a lower bound for the code overhead as defined in~\eqref{bound_field_overhead}, see~\cite{ICC2011}. The same lower bound holds here, as the lowest overhead is obtained when $w=g-1$, in which case the perpetual code becomes identical to \ac{RLNC}.

\eqref{bound_field_overhead} evaluates the expected overhead based on the probability that the rank increase at the receiver when a new coded symbol is received. This is a function of the generation size, $g$, the field size, $q$, and the rank at the receiver, $g'$.
For each of the indices where the decoder has already identified a pivot element, the coefficient in the incoming packet is reduced to zero by the decoder. In the best case, the remaining $g-g'$ elements can be considered as drawn at random from $\field{F}_q$. Hence the probability that these are all zero and the packet is linearly dependent is $1/q^{g-g'}$. The mean overhead is then calculated as the sum of the expected amount of overhead for the decoding of each packet, for all possible ranks of the decoder. Note that the overhead is primarily due to the last packets, and that it becomes negligible for high values of $q$.
\begin{align}
\alpha & \geq \sum_{g'=0}^{g-1} \left( \left(1- \frac{1}{q^{g-g'}} \right)^{-1}-1 \right) \notag \\ 
&  = \sum_{g'=0}^{g-1} \left( \frac{1}{q^{g-g'}-1} \right) \label{eq:bound_field_overhead}
\end{align}

%



For a symbol to be independent, either its pivot or one of the $w$ coefficient must hit a new pivot element. The pivot of the symbol is independent and hence the probability is $\frac{1}{g}$, but the $w$ elements depend on the pivot. The probability that one of these $w$ elements hits an uncovered pivot is $\frac{r'}{g}$ where $r'= [ 1,g-1 ]$. The expected number of tries to hit an unseen pivot is thus $\sum_{r'=g}^1 \left( \frac{r'}{g} \right) ^{-1} = g \cdot \sum_{r=0}^{g-1} \frac{1}{g-r}$. Thus the probability that one of the $w$ elements hits an unseen pivot can be expressed as $w / \left( g \cdot \sum_{r=0}^{g-1} \frac{1}{g-r} \right)$.
Then the probability that a symbol is covered when $x$ coded symbols have been received can be found as the probability that none of the $x$ coded symbols covers the symbol. In the worst case, decoding is possible when all $g$ pivots are covered.
%
\begin{align}
F_{X}(x) &\geq \left( 1- \left( 1- \left( \frac{1}{g}+ w / \left( g \cdot \sum_{r=0}^{g-1}  \frac{1}{g-r} \right) \right) \right) ^x \right) ^g \notag \\
\end{align}

The resulting \textit{cdf} can be used to calculate an upper bound for the code overhead by evaluating the corresponding \textit{survival function (sf)},
which defines the probability that there is an uncovered symbol after $x$ transmissions and thus additional transmissions are necessary.
%
\begin{align}
\beta & \leq \sum _{x=g}^\infty S_{X} (x) = \sum _{x=g}^\infty 1-F_{X} (x)\\
%
%
\alpha & \leq O \leq \alpha + \beta \notag 
\end{align}

\begin{figure}[htp]
\includegraphics[width=\columnwidth]{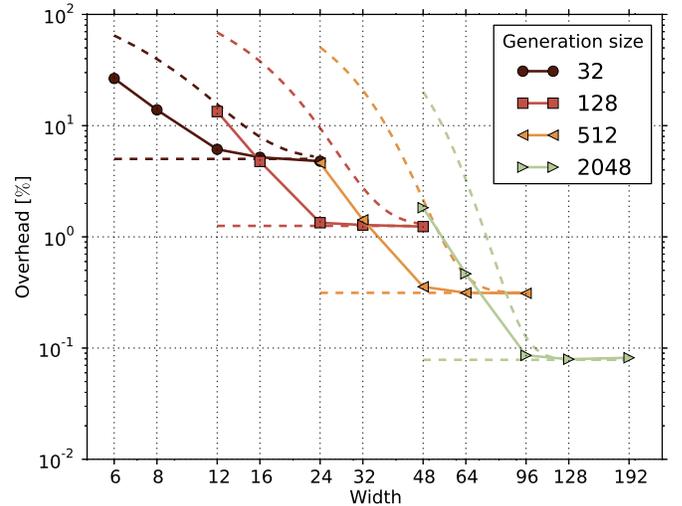}
\caption{Code overhead as a function of $g$ and $w$. The dotted lines denote the lower and upper bounds, respectively.}
\label{fig:overhead}
\end{figure}

In \figref{overhead}, the overhead for different generation sizes is plotted as a function of the code width (shown on the x-axis). The resulting overhead is given on the y-axis is. The dotted lines denote the upper and lower bounds, respectively.

For each generation size, the overhead decreases as the width increases until the width is sufficiently high and the overhead becomes indistinguishable from the lower bound. If the width is decreased below the sufficient level, the overhead increases significantly. Therefore, values of $w$ below this point should generally not be used. The bounds are loose for low values of $w$, but become tighter as $w$ increases. Thus the provided bounds are useful for identifying a value of $w$ that is sufficiently high.

%
We note that these results do not follow the overhead as a function of the density defined in~\cite{ICC2011}, which is similar to the width for this code. This is not surprising as the code investigated here is significantly less random compared to the sparse \ac{RLNC} considered in the reference.


\subsection{Complexity}

 We express the computational complexity using a compound metric called row \textit{multiplication-addition}, where a multiplication-addition is multiplying a row with a scalar and adding or subtracting it to or from another row. Here we only consider the binary field, therefore the multiplication scalar is always one and a row multiplication-addition is simply adding or subtracting a row to or from another row.

To \textit{encode} a single packet, the expected number of row operations is given by \eqref{encode_complexity}. We start with an empty vector and first add the chosen pivot row to it. For each of the following $w$ indices, the corresponding row is multiplied with a random element from $\mathbb{F}_q$ and added to the new row. The probability that a randomly drawn element from $\mathbb{F}_q$ is non-zero is $1-\frac{1}{q}$.
\begin{align}
1 + w \cdot (1-\frac{1}{q})
\label{eq:encode_complexity}
\end{align}

During \textit{on-the-fly} decoding, forward substitution is performed on the incoming symbol until a new pivot candidate is identified or the symbol is reduced to the $\vect{0}$ vector.
%
%
Forward substitution continues as long as the next element is non-zero, probability $1-\frac{1}{q}$, and has not already been identified as a pivot, probability $1-\frac{r}{g}$, where $r$ is the current rank of the decoding matrix. The expected number of row operations for a generation is found by summing over the reciprocal for all values of $r$, from which the expected number of operations per symbol is found by dividing with $g$.
\begin{align}
 \label{eq:delta0}
\delta_{\mathrm{fly}} \leq &  \frac{1}{g} \sum_{r=0}^{g-1} \left( \left( 1-\frac{1}{q}\right) \left( 1 - \frac{r}{g} \right)  \right) ^{-1} \notag \\
= &  \frac{q}{q-1} \cdot \frac{1}{g} \sum_{r=0}^{g-1} \frac{g}{g-r} \notag \\ 
= & \frac{q}{q-1} \sum_{r'=1}^{g} \frac{1}{r'}
\end{align}








For the upper bound for the \textit{final decoding}, we consider the worst case where most scalars are non-zero, see \figref{final_initial}.
The final forward stage on \figref{final_forward} can be considered in two steps.
First, the bottom $w$ rows are reduced, by substituting the top $g-w$ into them, so only the last $w$ elements are non-zero, hence \eqref{sub}.
Then the bottom $w$ rows are brought onto echelon form.
\eqref{inv} accounts for the forward substitution step in the bottom right $w \times w$ submatrix.
To include the probability that an element in $\mathbb{F}_q$ is equal to zero, we multiply with $\left( 1- \frac{1}{q}\right)$ and divide by $g$ to find the operations per packet, which is rewritten as $\left( \frac{q-1}{q\cdot g} \right)$.

\begin{align}
\delta_{\mathrm{forward1}} & \leq  \left( \frac{q-1}{q\cdot g} \right) \cdot (g - w) \cdot w  \label{eq:sub} \\
\delta_{\mathrm{forward2}} & \leq \left( \frac{q-1}{q\cdot g} \right) \cdot \sum _{i=1} ^{w-1} i \notag \\
&= \left( \frac{q-1}{q\cdot g} \right) \cdot \frac{w \cdot (w-1)}{2} \label{eq:inv}
\end{align}

To finalize the decoding, a similar procedure is performed, but this time upwards. Each of the $g-w$ bottom rows are substituted into the $w$ rows directly above them. 
 Thus the number of operations is exactly the same as in \eqref{sub} and \eqref{inv} and we obtain~\eqref{delta_2}.
\begin{align}
\delta & \leq \delta_{\mathrm{fly}} + 2 \delta_{\mathrm{forward1}} + 2 \delta_{\mathrm{forward2}} \notag \\
& = \frac{q}{q-1} \sum_{r'=1}^{g} \frac{1}{r'} + \left( \frac{q-1}{q\cdot g} \right) \left( w \cdot (2g-w-1 )  \right) \label{eq:delta_2}
\end{align}

\begin{figure}[htp]
\includegraphics[width=\columnwidth]{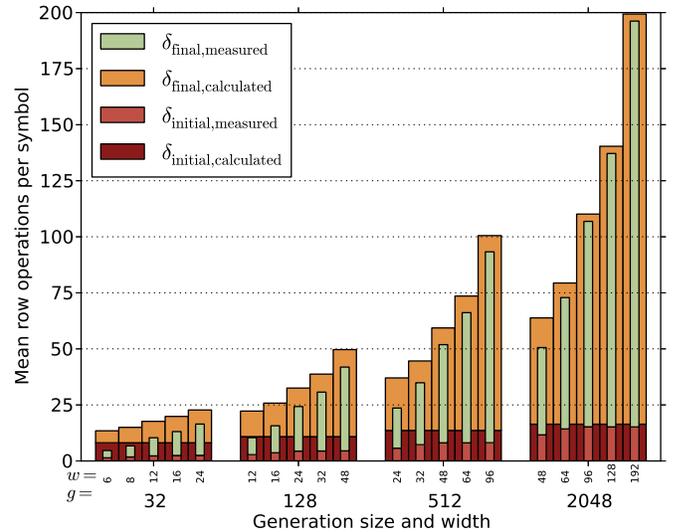}
\caption{Mean row operations per decoded symbol.}
\label{fig:operations_simple}
\end{figure}

\figref{operations_simple} shows the upper bound and measured number of row multiplication-additions performed to decode one generation, both during the \textit{on-the-fly} and \textit{final} decoding phase. The operations during the two phases are stacked to show the total number of row operations.

The analytical expressions for the \textit{on-the-fly} and \textit{final} decoding provide good bounds for the measured results, especially when $w$ is sufficiently high. For low values of $w$, the bound is less tight, but such settings should not be used when the code overhead is considered.

These values can be compared with traditional \ac{RLNC} where the expected number of operations to decode a packet is approximately $g/2$ for the binary case~\cite{NCpro2011A}. Thus the reduction in complexity compared to \ac{RLNC} grows as $g$ increases.

\subsection{Throughput}

A low complexity does not guarantee a low computational load and therefore we investigate the coding throughput.
This is due to the complexity introduced by the algorithms that determine how decoding is performed, the quality of their implementation and the bookkeeping they add. The architecture also affects the throughput due to cache misses, memory delay and throughput, which can also be influenced by the memory access pattern.


For each setting, data was encoded and subsequently decoded on the same machine. Each setting was run for a minimum of 30~minutes to reduce the deviation.

\begin{table}[htp]
\caption{Specifications of the test machine.}
\label{tab:dell}
\centering
\begin{tabular}{l l}
\toprule
Model    & Dell Optiplex 790 DT \\
\midrule
CPU      & Intel Core i7-2600 @ 3.40GHz, 8192 KB L2 cache \\
Memory   & 16GB DDR3 1333 MHz Dual channel \\
Chipset  & Intel Q65 Express \\
OS       & 64bit Debian Wheezy \\
Compiler & GNU G++ 4.6 \\
\bottomrule
\end{tabular}
\end{table}

To provide a comparison, we have performed benchmarks of our \ac{RLNC} implementation using the same values of $g$ as for the perpetual code. The encoding and decoding throughput's are listed in \tabref{throughput} along with the corresponding gains over \ac{RLNC}.

\begin{table}[htp]
\caption{Measured encoding and decoding throughput for \ac{RLNC} and the proposed perpetual code. Throughputs are reported for different generation sizes, $g$, and in the case of the perpetual code at different widths, $w$. The lowest tested value of $w$ where the perpetual code provides a similar code overhead as \ac{RLNC} is marked.}
\label{tab:throughput}
\centering
\renewcommand{\tabcolsep}{.15cm}
\begin{tabular}{c c r r r r r}
\toprule
\multirow{2}{*}{$g$} & \multirow{2}{*}{$w$} & Overhead & Encoding & Gain & Decoding & Gain\\
& & [packets] & [MB/s] & [\%] & [MB/s]  & [\%] \\
\midrule
   &  6 & 8.24 & 2883.91 & 156 & 2879.99 & 147 \\
   &  8 & 4.15 & 2512.14 & 123 & 2034.45 &  74 \\
\rowcolor{mygreen} \cellcolor{white} \multirow{2}{*}{
32}& 12 & 1.70 & 1915.58 &  70 & 1359.66 &  16 \\
   & 16 & 1.65 & 1506.90 &  34 & 1214.37 &   4 \\
   & 24 & 1.62 & 1080.19 &  -4 &  881.03 & -25 \\
 & RLNC & 1.61 & 1126.16 & -  & 1167.67 & - \\
\midrule
    & 12 & 17.06 & 1612.36 & 441 & 1209.09 & 326 \\
    & 16 &  6.05 & 1309.09 & 339 &  951.04 & 235 \\
\rowcolor{mygreen} \cellcolor{white} \multirow{2}{*}{
128}& 24 &  1.65 & 951.04 & 219 &  620.30 & 119 \\
    & 32 &  1.64 & 742.68 & 149 &  526.43 & 86 \\
    & 48 &  1.63 & 520.06 &  74 &  360.34 & 27 \\
   & RLNC & 1.61 & 298.28 & -   &  283.69 & - \\
\midrule
    & 24 & 24.33 & 747.29 & 921 & 490.50 & 655 \\
    & 32 &  7.02 & 612.69 & 737 & 392.39 & 504 \\
\rowcolor{mygreen} \cellcolor{white} \multirow{2}{*}{
512}& 48 &  1.68 & 449.39 & 514 & 273.00 & 320 \\
    & 64 &  1.65 & 354.48 & 384 & 228.40 & 251 \\
    & 96 &  1.63 & 249.25 & 241 & 167.27 & 157 \\
   & RLNC & 1.61 & 73.17 & -  & 64.99 & - \\
\midrule
     &  48 & 36.01 & 314.79 & 1656 & 203.11 & 1321\\
     &  64 &  9.03 & 263.36 & 1369 & 167.82 & 1074 \\
\rowcolor{mygreen}\cellcolor{white}\multirow{2}{*}{
2048}&  96 &  1.66 & 198.93 & 1010 & 129.93 &  809 \\
     & 128 &  1.64 & 160.38 &  795 & 99.90  & 599 \\
     & 192 &  1.62 & 115.00 &  541 & 66.81 & 368 \\
    & RLNC &  1.61 &  17.93 & - & 14.29 & -\\
\bottomrule
\end{tabular}
\end{table}

As expected, the throughput for both encoding and decoding decreases as $g$ and $w$ increase. The gain increases for higher values of $g$ which corresponds with the analytical results. The highest gain in encoding and decoding throughput is observed at the highest tested generation size of 2048, and approximately eleven and nine times that of \ac{RLNC} respectively. It should also be noted that using an excessively high $w$ should be avoided as it decreases the throughput without reducing the code overhead. Additionally, a higher $w$ increases the size of the coding vector representation, see \eqref{vector_bits}, which adds to the overall overhead.


To make a fair comparison with \ac{RLNC} we must consider both the overhead and the complexity / throughput simultaneously, as the performance of the perpetual code is a trade-off between overhead and speed. As the lower bounds are the same as for \ac{RLNC} we can never hope to achieve a lower code overhead. However, we can achieve a similar overhead but at lower computational complexity.
For this reason the throughput's for the perpetual approach is marked for the lowest value of $w$ where the code overhead is similar to \ac{RLNC}.



Ideally, all decoding should be performed \textit{on-the-fly} as this decreases the final decoding delay and distributes the processing load evenly. At the same time, decoding should be performed in such a way that \textit{fill-in} does not occur as this reduces the amount of work necessary to decode~\cite{Ingram_minimumdegree}. In our presented results, the ratio of operations performed during \textit{on-the-fly} phase is low, see~\figref{operations_simple}. Fortunately, the structure of the code makes it possible to perform something that can best be described as opportunistic backwards substitution. Our tests with this approach show that most of the decoding operations can be performed when symbols are received. However, this algorithm is more difficult to analyze and due to space constraints we have omitted it.

We note that the implementation does not take advantage of multiple cores. This could however be exploited by encoding multiple streams simultaneously or encoding simultaneously from different blocks of the same data set.

\subsection{Recoding}

In general, the performance of the proposed recoding approach will depend on the network topology, therefore general results are difficult to obtain. Instead, we consider the simplest multi-hop topology to provide fundamental insights into the recoding performance of the scheme.

Source $A$ transmits data to $R$ with some erasure probability. Both $R$ and $B$ send a single bit of feedback, namely when they have achieved full rank.
When a symbol arrives at $R$, it is forwarded to $B$ and to correct the $\epsilon_{RB}$ erasures on average $(1-\epsilon_{RB})^{-1}-1$ symbols are recoded at $R$ and transmitted to $B$. Initially, when too few symbols have been accumulated at $R$, it is pointless to attempt recoding, which is also the case for traditional \ac{RLNC}. Therefore, the remaining missing symbols will be re-encoded after $R$ has achieved full rank.

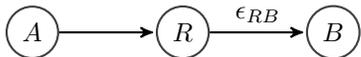
\begin{figure}[htp]
\centering

\begin{tikzpicture}[node distance=2cm,bend angle=45,auto]

  \tikzstyle{node}=[circle,thick,draw=black!75,minimum size=6mm]
  \tikzstyle{signal}=[->,>=stealth',thick,shorten >=1pt,rectangle,thick,minimum size=4mm]

    \node [node] (a)              {$A$};
    \node [node] (r) [right of=a] {$R$};
    \node [node] (b) [right of=r] {$B$};

    \path
    (a)  edge [signal] (r)
    (r)  edge  [signal] node {$\epsilon_{RB}$} (b);

\end{tikzpicture}
\caption{A simple multihop scenario.}
\label{fig:scenario}

\end{figure}

We define the parameter $1 < \mu < g$ which specifies the minimum number of symbols that should be combined to create a recoded symbol. In traditional \ac{RLNC}, typically all received symbols are combined at random, thus $\mu \approx r \cdot (1- \frac{1}{q})$.
\footnote{In general, the necessary amount of recoding depends on the network topology, and more specifically on the correlation of incoming links at nodes that receive recoded symbols.}
When $R$ combines $\mu$ symbols to create a recoded symbol, the probability that one or more symbols previously not seen at $B$ is included in the recoded symbol can be expressed as in \eqref{P_unseen}. 
\begin{align}
  P_{\mathrm{unseen}} \leq 1-(1-\epsilon_{RB})^\mu \label{eq:P_unseen}
\end{align}

For $\epsilon = 0.3$ \cite{MobiMedia2008} and $P_{\mathrm{unseen}} \approx 0.99$, $\mu =  12$.
%
%
%
%
%
When a new symbol is received at $R$, decoding is attempted and with some probability $\mu$ or more row operations are performed on the symbol in which case the resulting inserted symbol has been successfully \textit{passively} recoded. This probability, $P_\mathrm{passive}$, is found as the probability that a random sequence of $\mu$ symbols is non-zero, where the probability of each symbol can be calculated from $r$ and $g$.
\begin{align}
P_\mathrm{passive}(r,\mu)  &= \frac{r}{g} \cdot \frac{r-1}{g-1} \cdot \hdots \cdot \frac{r - (\mu - 1) }{g - (\mu-1)} \notag \\
&= \prod_{i=0}^{\mu-1} \frac{r-i}{g-i}
\end{align}

Otherwise \textit{active} recoding becomes necessary, and is possible if $\mu$ pivot elements have been identified in some range of size $\Delta$. The symbols in this range are combined at random and the resulting symbol will have a width $w' \leq w + \Delta$.
The maximal width accepted is denoted $w'_{\mathrm{max}} = w + \Delta_\mathrm{max}$. Here we assume that $\Delta_\mathrm{max} = 2 \cdot \mu$ in order to permit some freedom during recoding. 
%

Consider a range of size $\Delta$, the number of ranges where at least one pivot element is zero is defined as $\sum_{j=0}^{\mu-1} \left( {\Delta \choose j} {{g-\Delta} \choose {r-j}} \right)$. From this and the total number of combinations ${g \choose r}$, the probability that a range contains $\mu$ or more pivots can be found. As there are $g$ such ranges we can find the probability that at least one range is suitable, see \eqref{P_active}.
\begin{align}
P_\mathrm{active} (r,\mu, \Delta) = 1 - \left( \sum_{j=0}^{\mu-1} \left( {\Delta \choose j} {{g-\Delta} \choose {r-j}} \right) / {g \choose r}  \right) ^g \label{eq:P_active}
\end{align}




The rest of recoding is performed as re-encoding after $B$ has obtained full rank.
\begin{align}
P_\mathrm{re-encode}(r,\mu,\Delta _\mathrm{max}) = 1 - P_\mathrm{passive}(r,\mu) - \notag \\ \sum_{\Delta = \mu}^{\Delta_{\mathrm{max}}} P_{\mathrm{active}} (r, \mu, \Delta)
\end{align}

\begin{figure}[htp]
\includegraphics[width=\columnwidth]{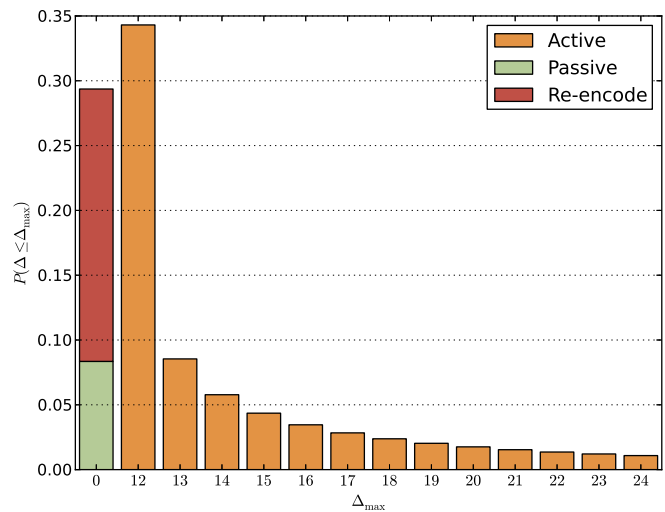}
\caption{Distribution of $\Delta$ for generated recoded symbols for $g=512$, $w=48$, $\epsilon_{RA}=0.3$, $\mu = 12$, and $\Delta_{\mathrm{max}} = 2 \cdot \mu$.}
\label{fig:recode_probability}
\end{figure}

Finally, we sum over $r = [1,g]$ for $P_\mathrm{passive}$, $P_\mathrm{active}$, and $P_\mathrm{re-encode}$ to obtain the distribution of the recoded output symbols. 
Passive recoding is preferred over active recoding, and active recoding over a smaller range is desirable. The resulting distribution is illustrated in \figref{recode_probability}. The x-axis denotes the maximum $\Delta$ and the y-axis denotes the expected ratio of recoded packets. Probabilities for passive recoding, active recoding, and re-encoding are shown.

For 30\% of the symbols, whereof 20\% is re-encoded, $w' = w$ and the generated symbols are indistinguishable from symbols encoded at a source.
For a third of the generated symbols $w'=48+12=60$ is slightly larger than $w=40$. The remaining generated symbols have $w'$ in the range [61,72].
This demonstrates that most times a recoded symbol can be generated and that the expected $\Delta$ is low. This is important in order to insure a low delay at $R$ and fast decoding at $B$.

For lower $g$, recoding becomes more difficult as $w'$ approaches $g$. However, more advanced decoding algorithms can be employed to reduce the added $\Delta$, unfortunately space does not permit the inclusion of these.









\section{Conclusion} \label{sec:conclusion}

In this paper, we presented our initial findings on \textit{perpetual codes} which are suitable for \ac{RLNC}. We described how encoding, decoding, and recoding can be performed and listed the necessary algorithms. We provided initial analysis of the code performance in terms of overhead and complexity. The analytical results were compared with measurements obtained from our C++ implementation from which we also obtained coding throughput measurements. 

The analysis and the tests showed that the proposed approach can obtain a coding overhead similar to \ac{RLNC}, but at a much lower computational cost. For all tested settings resulting in a code overhead similar to that of \ac{RLNC}, the proposed approach led to improved encoding and decoding throughput. For the highest tested generation size of 2048, the decoding throughput was almost one order of magnitude higher than that of \ac{RLNC}. Additionally, the approach provides an easily adjustable parameter that allows for a trade-off between coding complexity and code overhead.

Throughout this work, we have compared the proposed perpetual code with \ac{RLNC} and \ac{SRLNC} and not with other rateless \ac{FEC} codes. The reason is that \ac{NC} codes allow for recoding where traditional \ac{FEC} codes do not, thus they are less  suitable for cooperative networks. Compared to \ac{RLNC} and \ac{SRLNC}, the perpetual code provides the following benefits depending on the chosen values of $g$ and $w$:


\begin{enumerate}
\item Faster encoding, recoding, and decoding.
\item Sparsity is retained when recoding.
\item Small coding vector representation.
\item Simple decoding algorithms.
\end{enumerate}

As the code is sparse, fast encoding is trivial. Fast decoding is possible due to the structure of the code that helps to avoid \textit{fill-in} during decoding. Recoding can be performed fast by using the suggested \textit{passive} recoding approach. We note that this trick can also be employed for other \ac{NC} codes.

The structure and density of the coded packets can be retained if recoding is performed more carefully than what has been proposed for standard \ac{RLNC}, we have denoted this \textit{active} recoding. We note that doing so limits the degrees of freedom when recoding, but we believe that our proposed \textit{passive plus active} recoding presents a good trade-off between these two approaches.

As the location of the non-zero elements are well defined, it is trivial to create compact representations of the coding vectors. We believe that this is important as the commonly used assumption of a pseudo random function can be used to compress the coding vector cannot be used if recoding is to be supported~\cite{ICC2011}. Thus the size of the coding vector must be included in the total overhead.

The presented decoding algorithms are slightly more complicated than for standard \ac{RLNC}. However, due to the sparsity and structure of the coded symbols, it is possible to eliminate many of the inspections that are necessary when inverting the coding matrix.


We have only considered the \textit{random} encoding mode, meaning that the pivot element is always drawn at random and independently of the previous pivots, see \tabref{modes}. This corresponds to the worst-case where the channel is extremely lossy and thus systematic approaches are of no benefit. In cases where the erasure probability is low or moderate, a systematic or sequential  mode could be used which would decrease the code overhead and in particular the decoding complexity.




For the future, more rigorous analysis of the code overhead is necessary, especially for the case where low values of $w$ are used. Such analysis would be useful when more advanced variants of the perpetual code are studied.
For our implementation, we plan to perform tests using higher field sizes 
 and perform benchmarks on mobile devices. This could help to understand how to choose optimal parameters and to demonstrate the validity of the proposed solution on mobile phones and tablets.

\section*{Acknowledgment}
This work was partially financed by the CONE project (Grant No. 09-066549/FTP) granted by Danish Ministry of Science, Technology and Innovation.

\bibliographystyle{IEEEtran}
\bibliography{bibtex}

\begin{thebibliography}{10}
\providecommand{\url}[1]{#1}
\csname url@samestyle\endcsname
\providecommand{\newblock}{\relax}
\providecommand{\bibinfo}[2]{#2}
\providecommand{\BIBentrySTDinterwordspacing}{\spaceskip=0pt\relax}
\providecommand{\BIBentryALTinterwordstretchfactor}{4}
\providecommand{\BIBentryALTinterwordspacing}{\spaceskip=\fontdimen2\font plus
\BIBentryALTinterwordstretchfactor\fontdimen3\font minus
  \fontdimen4\font\relax}
\providecommand{\BIBforeignlanguage}[2]{{%
\expandafter\ifx\csname l@#1\endcsname\relax
\typeout{** WARNING: IEEEtran.bst: No hyphenation pattern has been}%
\typeout{** loaded for the language `#1'. Using the pattern for}%
\typeout{** the default language instead.}%
\else
\language=\csname l@#1\endcsname
\fi
#2}}
\providecommand{\BIBdecl}{\relax}
\BIBdecl

\bibitem{ahlswede}
R.~Ahlswede, N.~Cai, S.-Y.~R. Li, and R.~W. Yeung, ``Network information
  flow,'' \emph{IEEE Transactions on Information Theory}, vol.~46, no.~4, pp.
  1204--1216, 2000.

\bibitem{rlnc}
T.~Ho, R.~Koetter, M.~M\'{e}dard, D.~Karger, and M.~ros, ``The benefits of
  coding over routing in a randomized setting,'' in \emph{Proceedings of the
  IEEE International Symposium on Information Theory, ISIT '03}, June 29 - July
  4 2003.

\bibitem{Yang:2011:LFT:2093698.2093815}
S.~Yang and R.~W. Yeung, ``Large file transmission in network-coded networks
  with packet loss: a performance perspective,'' in \emph{Proceedings of the
  4th International Symposium on Applied Sciences in Biomedical and
  Communication Technologies}, ser. ISABEL '11.\hskip 1em plus 0.5em minus
  0.4em\relax Barcelona, Spain: ACM, 2011, pp. 117:1--117:5.

\bibitem{NCRealWorld}
J.~Heide, M.~V. Pedersen, F.~H. Fitzek, and T.~Larsen, \emph{Network Coding in
  the Real World}.\hskip 1em plus 0.5em minus 0.4em\relax Academic Press,
  October 2011, ch.~4, pp. 87--114.

\bibitem{perpetual}
\BIBentryALTinterwordspacing
P.~Maymounkov. (2006) Perpetual codes: cache-friendly coding. Unpublished
  draft, retieved 2nd of September 2011. [Online]. Available:
  \url{http://pdos.csail.mit.edu/~petar/papers/maymounkov-perpetual.ps}
\BIBentrySTDinterwordspacing

\bibitem{Ingram_minimumdegree}
\BIBentryALTinterwordspacing
S.~Ingram. (2006) Minimum degree reordering algorithms: A tutorial. Retrieved
  March 2010. [Online]. Available:
  \url{www.cs.ubc.ca/~sfingram/cs517_final.pdf}
\BIBentrySTDinterwordspacing

\bibitem{Chou03practicalnetwork}
P.~A. Chou, Y.~Wu, and K.~Jain, ``Practical network coding,'' \emph{Proceedings
  of the annual Allerton conference on communication control and computing},
  vol.~4, pp. 40--49, 2003.

\bibitem{Neubauer:2007}
A.~Neubauer, J.~Freudenberger, and V.~Kuhn, \emph{Coding Theory: Algorithms,
  Architectures and Applications}.\hskip 1em plus 0.5em minus 0.4em\relax
  Wiley-Interscience, 2007, appendix A provides an in depth review of algebraic
  structures, including finite fields.

\bibitem{primer}
C.~Fragouli, J.~Boudec, and J.~Widmer, ``Network coding: an instant primer,''
  \emph{SIGCOMM Comput. Commun. Rev.}, vol.~36, no.~1, pp. 63--68, 2006.

\bibitem{chunked_codes}
P.~Maymounkov, N.~J.~A. Harvey, and D.~S. Lun, ``{Methods for Efficient Network
  Coding},'' \emph{44th Allerton Annual Conference}, 2006.

\bibitem{yao}
Y.~Li, E.~Soljanin, and P.~Spasojević~and, ``Collecting coded coupons over
  overlapping generations,'' in \emph{Network Coding (NetCod), 2010 IEEE
  International Symposium on}, june 2010, pp. 1 --6.

\bibitem{5634159}
O.~Trullols-Cruces, J.~Barcelo-Ordinas, and M.~Fiore, ``Exact decoding
  probability under random linear network coding,'' \emph{Communications
  Letters, IEEE}, vol.~15, no.~1, pp. 67 --69, january 2011.

\bibitem{6188495}
X.~Zhao, ``Notes on "exact decoding probability under random linear network
  coding",'' \emph{Communications Letters, IEEE}, vol.~16, no.~5, pp. 720
  --721, may 2012.

\bibitem{ICC2009}
J.~Heide, M.~V. Pedersen, F.~H. Fitzek, and T.~Larsen, ``Network coding for
  mobile devices - systematic binary random rateless codes,'' in \emph{The IEEE
  International Conference on Communications (ICC)}, Dresden, Germany, 14-18
  June 2009.

\bibitem{shojania-parrallel}
H.~Shojania and B.~Li, ``Parallelized progressive network coding with hardware
  acceleration,'' in \emph{Quality of Service, 2007 Fifteenth IEEE
  International Workshop on}, June 2007, pp. 47--55.

\bibitem{jcn2008}
J.~Heide, M.~V. Pedersen, F.~H. Fitzek, and T.~Larsen, ``Cautious view on
  network coding - from theory to practice,'' \emph{Journal of Communications
  and Networks (JCN)}, vol.~10, no.~4, pp. 403--411, December 2008.

\bibitem{chinese-gfx}
X.~Chu, K.~Zhao, and M.~Wang, ``Massively parallel network coding on gpus,'' in
  \emph{Performance, Computing and Communications Conference, 2008. IPCCC 2008.
  IEEE International}, December 2008, pp. 144--151.

\bibitem{shojania-iphone}
H.~Shojania and B.~Li, ``Random network coding on the iphone: fact or
  fiction?'' in \emph{NOSSDAV '09: Proceedings of the 18th international
  workshop on Network and operating systems support for digital audio and
  video}.\hskip 1em plus 0.5em minus 0.4em\relax ACM, June 2009, pp. 37--42.

\bibitem{binary}
H.~feng (Francis)~Lu, ``Binary linear network codes,'' in \emph{IEEE
  Information Theory Workshop on Information Theory for Wireless Networks},
  July 2007.

\bibitem{sparse}
X.~Li, W.~H. Mow, and F.-L. Tsang, ``Singularity probability analysis for
  sparse random linear network coding.'' in \emph{IEEE International Conference
  on Communications (ICC)}, June 2011.

\bibitem{ICC2011}
J.~Heide, M.~V. Pedersen, F.~H. Fitzek, and M.~M\'{e}dard, ``On code parameters
  and coding vector representation for practical rlnc,'' in \emph{IEEE
  International Conference on Communications (ICC) - Communication Theory
  Symposium}, Kyoto, Japan, jun 2011.

\bibitem{NCpro2011A}
J.~Heide, M.~V. Pedersen, and F.~H. Fitzek, ``Decoding algorithms for random
  linear network codes,'' in \emph{IFIP International Conferences on Networking
  - Workshop on Network Coding Applications and Protocols (NC-Pro)}, ser.
  Lecture Notes in Computer Science, vol. 6827, Valencia, Spain, may 2011, pp.
  129--137.

\bibitem{citeulike:10205307}
E.~Erez and M.~Feder, ``Convolutional network codes,'' in \emph{Information
  Theory, 2004. ISIT 2004. Proceedings. International Symposium on}.\hskip 1em
  plus 0.5em minus 0.4em\relax IEEE, Jun. 2004, pp. 146+.

\bibitem{citeulike:10205303}
S.~Y.~R. Li and R.~W. Yeung, ``On convolutional network coding,'' in
  \emph{Information Theory, 2006 IEEE International Symposium on}.\hskip 1em
  plus 0.5em minus 0.4em\relax IEEE, Jul. 2006, pp. 1743--1747.

\bibitem{raptor}
A.~Shokrollahi, ``{Raptor codes},'' \emph{IEEE Transactions on Information
  Theory}, vol.~52, no.~6, pp. 2551--2567, Jun. 2006.

\bibitem{Fragouli04aconnection}
C.~Fragouli and E.~Soljanin, ``A connection between network coding and
  convolutional codes,'' in \emph{IEEE International Conference on
  Communications (ICC)}, 2004.

\bibitem{equivalent}
S.~Jaggi, M.~Effros, T.~Ho, and M.~Médard, ``On linear network coding,'' in
  \emph{42st Annu. Allerton Conf. Communication Control and Computing}, 2004.

\bibitem{silva}
D.~Silva, W.~Zeng, and F.~Kschischang, ``Sparse network coding with overlapping
  classes,'' in \emph{Network Coding, Theory, and Applications, 2009. NetCod
  '09. Workshop on}, june 2009, pp. 74 --79.

\bibitem{DBLP:journals/corr/abs-1011-3498}
Y.~Li, E.~Soljanin, and P.~Spasojevic, ``Effects of the generation size and
  overlap on throughput and complexity in randomized linear network coding,''
  \emph{CoRR}, vol. abs/1011.3498, 2010.

\bibitem{kodo_perpetual}
\BIBentryALTinterwordspacing
(2012) Perpetual code implementation. Currently not publically available, but
  will be at time of publication. [Online]. Available:
  \url{https://github.com/steinwurf/kodo/tree/master/kodo/perpetual}
\BIBentrySTDinterwordspacing

\bibitem{MobiMedia2008}
J.~Heide, M.~V. Pedersen, F.~H. Fitzek, T.~V. Kozlova, and T.~Larsen, ``Know
  your neighbour: Packet loss correlation in ieee 802.11b/g multicast,'' in
  \emph{The 4th International Mobile Multimedia Communications Conference
  (MobiMedia '08)}, Oulu, Finland, July 7-9 2008.

\end{thebibliography}

\begin{biography}[{\includegraphics[width=1in,height=1.25in,clip,keepaspectratio]{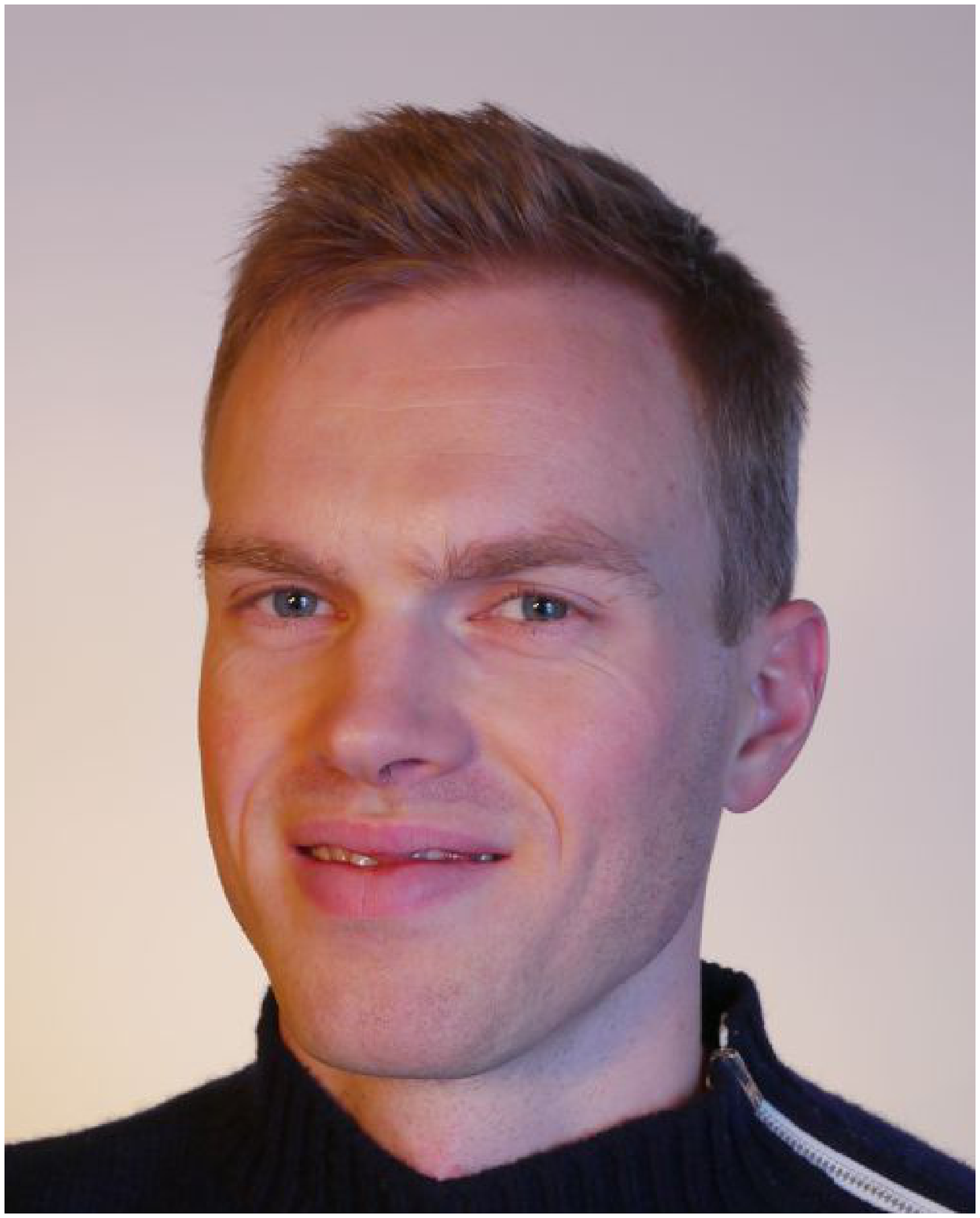}}]{Janus Heide}
is a postdoctoral researcher at Aalborg University. He received his M.Sc in electrical engineering with specialization Wireless Communication Engineering and his Ph.D. from Aalborg University, Denmark, in 2009 and 2012 respectively. From July 2007 he has been working in the Mobile Devices research group at Aalborg University. His main research interests are protocol analysis and design, network coding, wireless and cooperative communication, and data distribution in meshed networks.
\end{biography}
\begin{biography}[{\includegraphics[width=1in,height=1.25in,clip,keepaspectratio]{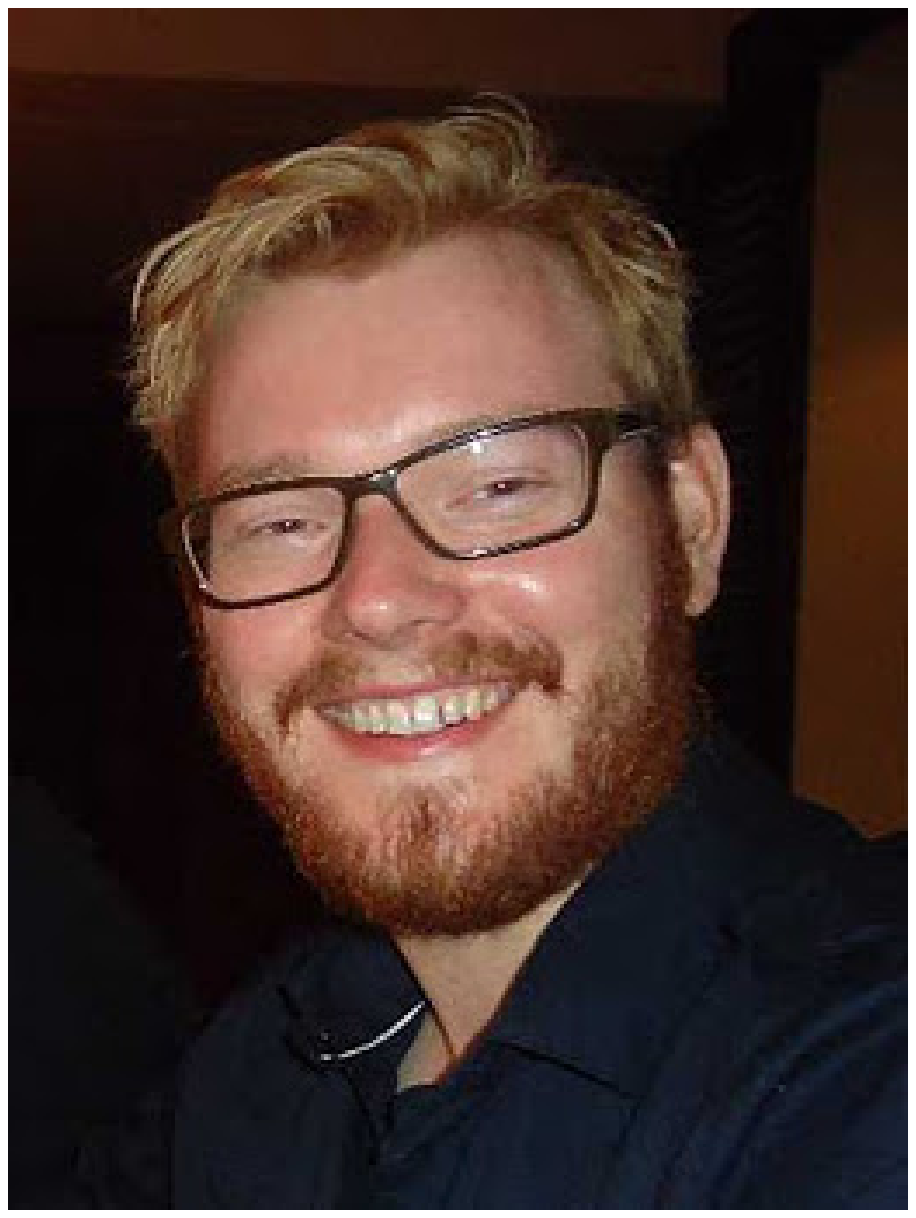}}]{Morten V. Pedersen}
is a postdoctoral researcher at Aalborg University. He received his M.Sc. in wireless communication in 2009, and his Ph.D. in 2012 both from Aalborg University, Denmark. From 2006 he has been working in the Mobile Devices research group at Aalborg University. In 2010 he was appointed Forum Nokia Champion. His main research interests are mobile programming, cooperative communication, network coding and network performance evaluation.
\end{biography}
\begin{biography}[{\includegraphics[width=1in,height=1.25in,clip,keepaspectratio]{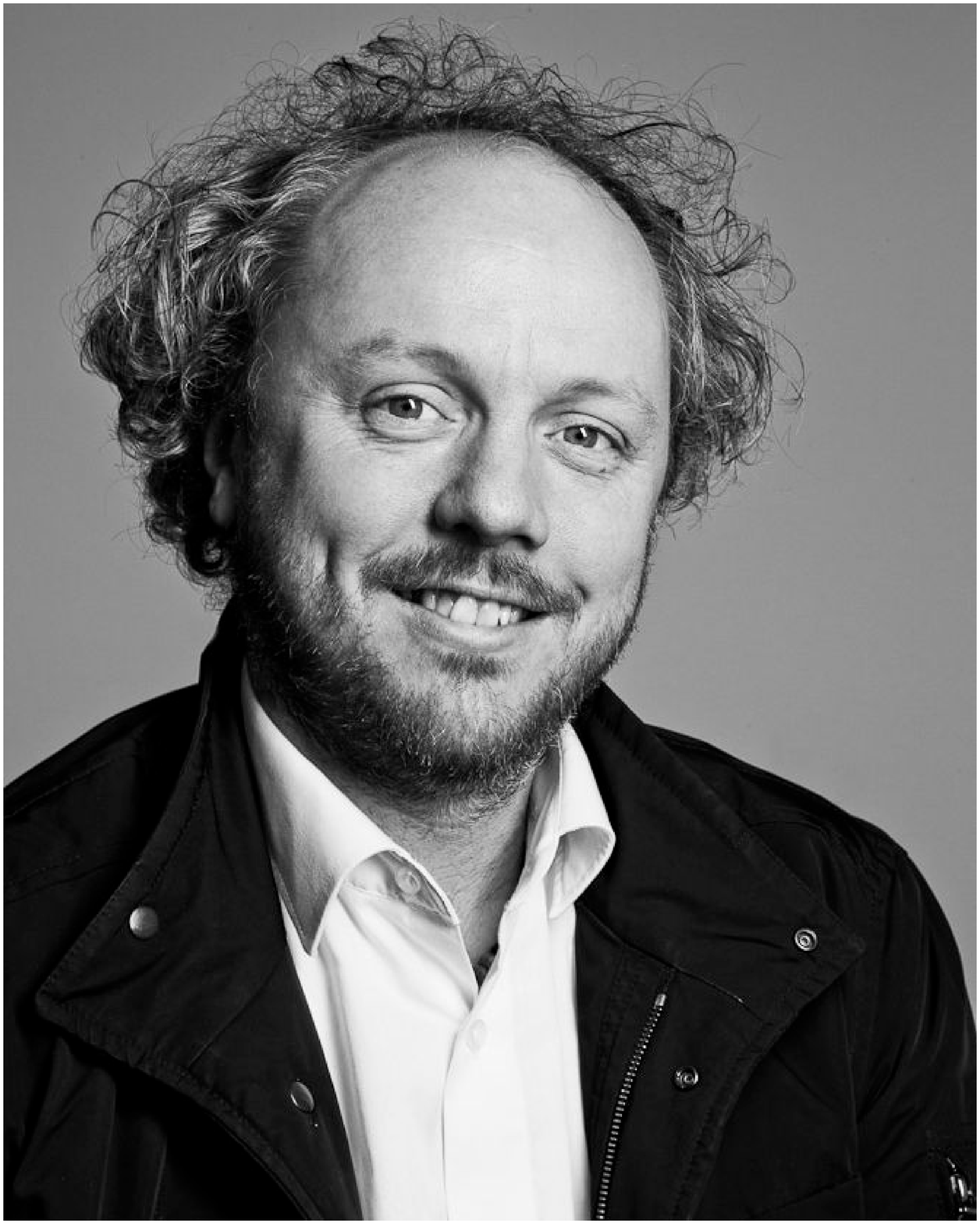}}]{Frank H.P. Fitzek}
is a Professor in the department of Electronic Systems, Aalborg University, Denmark heading the Mobile Device group. 
In 2005 he won the YRP award and received the Young Elite Researcher Award of Denmark. He was selected to receive the NOKIA Champion Award five times in a row from 2007 to 2011. In 2011 he received the SAPERE AUDE research grant from the Danish government and in 2012 he received the Vodafone Innovation price. His current research interests are in the areas of wireless and mobile communication networks, mobile
phone programming, cross layer as well as energy efficient protocol design and cooperative networking.
\end{biography}
\begin{biography}[{\includegraphics[width=1in,height=1.25in,trim=20 0 20 0, clip,keepaspectratio]{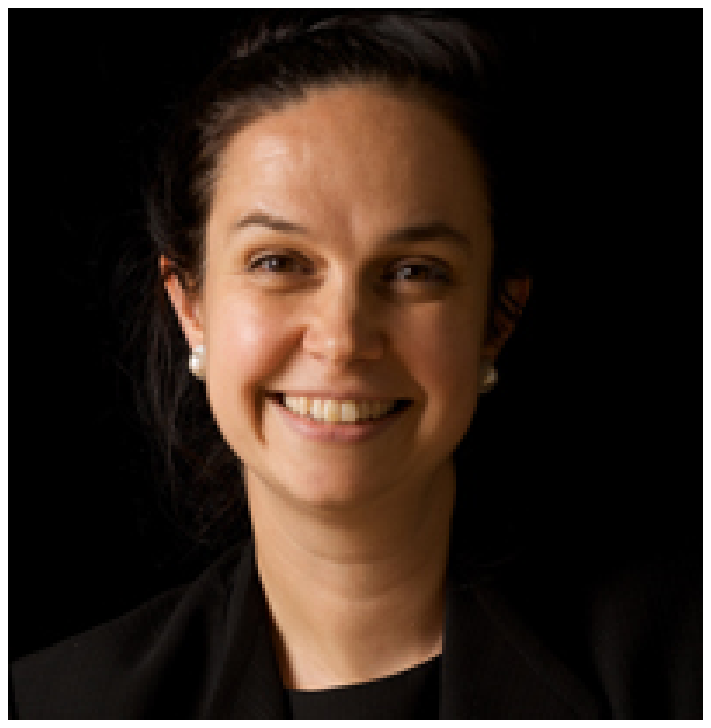}}]{Muriel M{\'e}dard}
is a Professor in the Electrical Engineering and Computer Science at MIT. Professor Médard received B.S. degrees in EECS and in Mathematics in 1989, a B.S. degree in Humanities in 1990, a M.S. degree in EE 1991, and a Sc D. degree in EE in 1995, all from the Massachusetts Institute of Technology (MIT), Cambridge. She has served as editor for numerous journals and on the board of Governors of the IEEE Information Theory Society as well as serving as the first Vice President in 2011 and the President in 2012.
 Her research interests are in the areas of network coding and reliable communications, particularly for optical and wireless networks.
%
\end{biography}

\end{document}